\documentclass[a4paper,usenatbib,useAMS]{mnras}

\usepackage[T1]{fontenc}
\usepackage{ae,aecompl}
\usepackage{amsmath}
\usepackage{amsfonts}
\usepackage{amssymb}
\usepackage{gensymb}
\usepackage{graphicx}
\usepackage{mathptmx}
\usepackage{multirow}
\usepackage{subfig}
\usepackage{tabularx}
\usepackage{threeparttable}
\usepackage{pdflscape}

\title[An AMI-LA survey of the SuperCLASS super-cluster]{AMI-LA Observations of the SuperCLASS Super-cluster}

\label{firstpage}

\author[C.~J.~Riseley et al.]{C.~J.~Riseley$^{1}$\thanks{Corresponding author email: \href{mailto:chris.riseley@csiro.au}{chris.riseley@csiro.au}}, K.~J.~B.~Grainge$^2$, Y.~C.~Perrott$^3$, A.~M.~M.~Scaife$^2$,  R.~A.~Battye$^2$, 
\newauthor R.~J.~Beswick$^2$, M.~Birkinshaw$^4$, M.~L.~Brown$^2$, C.~M.~Casey$^5$, C.~Demetroullas$^{2,6}$,
\newauthor C.~A.~Hales$^{7}$\thanks{Marie Sk\l{}odowska-Curie Fellow.}, I.~Harrison$^2$, C.-L. Hung$^{5}$, N.~J.~Jackson$^2$, T.~Muxlow$^2$, B.~Watson$^2$,
\newauthor  T.~M.~Cantwell$^2$, S.~H.~Carey$^3$, P.~J.~Elwood$^3$, J.~Hickish$^{3,8}$,  T.~Z.~Jin$^3$, N.~Razavi-Ghods$^3$, 
\newauthor  P.~F.~Scott$^3$, D.~J.~Titterington$^3$
 \vspace{0.03in}\\
$^1$ CSIRO Astronomy \& Space Science, 26 Dick Perry Avenue, Kensington, WA 6151, Australia \\
$^2$ Jodrell Bank Centre for Astrophysics, Alan Turing Building, School of Physics and Astronomy, The University of Manchester, Oxford Road, Manchester, \\
M13 9PL, U.K.\\
$^3$ Astrophysics Group, Cavendish Laboratory, 19 J.~J.~Thomson Avenue, Cambridge, CB3 0HE, UK \\
$^4$ H.~H.~Wills Physics Laboratory, University of Bristol, Tyndall Avenue, Bristol BS8 1TL, U.K. \\
$^5$ Department of Astronomy, The University of Texas at Austin, 2515 Speedway Blvd Stop C1400, Austin, TX 78712, USA. \\
$^6$ The Cyprus Institute, 20 Konstantinou Kavafi Street, 2121, Aglantzia, Nicosia, Cyprus. \\
$^7$ School of Mathematics, Statistics and Physics, Newcastle University, Newcastle upon Tyne NE1 7RU, UK \\
$^8$ Radio Astronomy Laboratory, University of California, Berkeley, CA 94720, USA \\
}

\date{Accepted 2017 November 29; received 2017 November 29; in original form 2017 October 22}

\pubyear{2017}

\begin{document}
%------------------------------------------------------------------------------%
\maketitle

\begin{abstract}
We present a deep survey of the SuperCLASS super-cluster -- a region of sky known to contain five Abell clusters at redshift $z\sim0.2$ -- performed using the Arcminute Microkelvin Imager (AMI) Large Array (LA) at 15.5~GHz. Our survey covers an area of approximately 0.9 square degrees. We achieve a nominal sensitivity of $32.0~\umu$Jy beam$^{-1}$ toward the field centre, finding 80 sources above a $5\sigma$ threshold. We derive the radio colour-colour distribution for sources common to three surveys that cover the field and identify three sources with strongly curved spectra -- a high-frequency-peaked source and two GHz-peaked-spectrum sources. The differential source count (i) agrees well with previous deep radio source count, (ii) exhibits no evidence of an emerging population of star-forming galaxies, down to a limit of 0.24~mJy, and (iii) disagrees with some models of the 15~GHz source population. However, our source count is in agreement with recent work that provides an analytical correction to the source count from the SKADS Simulated Sky, supporting the suggestion that this discrepancy is caused by an abundance of flat-spectrum galaxy cores as-yet not included in source population models.

\end{abstract}


\defcitealias{Shimwell2013}{AMI Consortium: Shimwell et al. 2013}
\defcitealias{franzen2011}{AMI Consortium: Franzen et al. 2011}
\defcitealias{zwart2008}{AMI Consortium: Zwart et al. 2008}

\section{Introduction}
Whilst low-frequency radio astronomy has experienced a renaissance in recent years, the high-frequency radio sky $(\nu \gtrsim 10~{\rm{GHz}})$ has remained relatively unexplored. The 9th Cambridge radio survey (9C; \citealt{waldram2003,waldram2010}) achieved a completeness level of 5.5 mJy across 29 square degrees of sky with the Ryle Telescope at 15~GHz; the wider survey covered 520 square degrees to a limit of approximately 25 mJy. More recently, the 10th Cambridge survey (10C; e.g \citetalias{franzen2011}) used the Ryle Telescope's successor, the Arcminute Microkelvin Imager (AMI; \citetalias{zwart2008}) to survey a number of fields totaling approximately 27 square degrees to a typical detection threshold of $\sim1$ mJy at 15.7~GHz \citep{franzen2011}. 

Smaller deep regions of the 10C survey cover $\sim12$ square degrees to $\sim0.5$ mJy sensitivity. Recently, \cite{whittam2016} present further deep observations of selected fields from the 10C survey area with AMI, achieving a best sensitivity of $16~\umu$Jy beam$^{-1}$, from which they derive the differential source count to a flux limit of $100~\umu$Jy. 

In the Southern hemisphere, the Australia Telescope 20GHz survey (AT20G; \citealt{murphy2010}) covered the entire sky at declination $\delta < 0\degree$ to a limit of 40 mJy; the survey is 93 per cent complete above 100 mJy. Additionally, \cite{franzen2014} present deeper observations of the \emph{Chandra} Deep Field South and SDSS Stripe 82 to a 90 per cent completeness level of 2.5 mJy. 

Whilst the source population at lower frequencies is well-constrained - steep-spectrum objects dominate at higher flux densities, with populations of star-forming galaxies becoming increasingly important at low flux densities, and dominating below $\sim1$ mJy at 1.4~GHz - the high-frequency source population remains less well understood. Numerous attempts have been made to model this population, typically based on extrapolation from lower frequencies. Older evolutionary models of radio source populations (e.g. \citealt{1990MNRAS.247...19D,1998MNRAS.297..117T,1999MNRAS.304..160J}) provide successful fits to data below 8~GHz to flux densities of the order of a few mJy. 

More recently, \cite{zotti2005} derive models of the radio source population above 5~GHz (this was later complemented at low frequencies by \citealt{massardi2010}) considering flat-spectrum sources (such as BL Lac objects and flat-spectrum radio quasars), steep-spectrum sources and star-forming galaxies. We refer the reader to \cite{zotti2010} for a review of source population models across a wide range of frequencies; see also \cite{padovani2016} for a recent comprehensive review of the current state of knowledge about radio source populations down to very faint flux densities.

\cite{wilman2008} developed a semi-empirical simulation of the extragalactic radio continuum sky for the Square Kilomtetre Array (SKA) - known as the SKADS Simulated Sky, or S$^3$. This model includes five separate source populations: radio-quiet AGN, radio-loud AGN - both Fanaroff-Riley Type I and Type II (FRI and FRII, respectively) sources - and star-forming galaxies (both quiescent and starburst). From this simulation the authors extract a simulated catalogue of $\sim320$ million radio sources at frequencies between 151~MHz and 18~GHz.

However, S$^3$ fails to replicate the observed spectral index distribution for sources in the 10C and AT20G catalogues (respectively \citealt{whittam2013} and \citealt{mahony2011}) and also the observed source count from the 9C/10C surveys \citep{whittam2016}. These observations indicate the presence of a substantial population of flat-spectrum sources that are not accounted for by simulations or extrapolations from lower-frequency models \citep{whittam2013,whittam2016}. 

Recent very high resolution observations at 1.4~GHz also support this, with the detection of a highly core-dominated AGN population\footnote{Sometimes referred to in the literature as `FR0' sources \citep[e.g.][]{Baldi2015}} which makes up a significant proportion of the sub-mJy source population \citep{Baldi2015,HerreraRuiz2017}. Very recently, \cite{Whittam2017} have shown that an additional flat-spectrum component added to the population of FRI sources in S$^3$ reproduces both the 9C/10C source count and the observed spectral index distribution between 15 and 1.4~GHz. 

In this work, we present observations of a galaxy super-cluster with the AMI Large Array (LA) at 15.5~GHz. This work was performed as part of the Super-Cluster Assisted Shear Survey (SuperCLASS). The SuperCLASS project is an e-MERLIN legacy survey at L-band, whose principal goal is detecting the effect of cosmic shear in the radio regime. We refer the reader to Battye et al. (in prep) for more details of the SuperCLASS project. Additionally, we have previously produced the deepest 325~MHz survey to date (nominal sensitivity of $34~\umu$Jy beam$^{-1}$) with the GMRT as part of this project \citep{riseley2016} covering an area of 6.5 square degrees around the super-cluster. Some observational properties of clusters in the SuperCLASS super-cluster are presented in Table \ref{tab:clusters}.

The remainder of this paper is divided as follows: we discuss our observations and data reduction methodology in \S\ref{sec:obs}. We present our results in \S\ref{sec:res}, including a sample from our source catalogue; we verify the catalogue and analyse the statistical properties in \S\ref{sec:an}, including an initial investigation into multi-wavelength properties of sources in our catalogue. We derive the source count distribution from our catalogue, as well as evaluate the various sources of bias, in \S\ref{sec:src}. Finally, we draw our conclusions in \S\ref{sec:conc}. All errors are quoted to $1\sigma$. We adopt the spectral index convention that $S \propto \nu^{\alpha}$. We assume a concordance cosmology of H$_0 = 73~\rm{km}~\rm{s}^{-1}$~Mpc$^{-1}$, $\Omega_{\rm{m}} = 0.27$, $\Omega_{\rm{\Lambda}} = 0.73$. At a redshift of $z = 0.2$, representative of the constituent clusters of the SuperCLASS supercluster, an angular size of 1 arcsecond corresponds to a physical size of 3.2 kpc.

\begin{table}
\begin{center}
\caption{Properties of galaxy clusters constituting the SuperCLASS supercluster.}
\label{tab:clusters}
\begin{threeparttable}
\begin{tabular}{cccccc}
\hline
 & & & \\ 
Name & RA & Dec & $z$ & $L_{\rm{x}}$ (0.1-2.4 keV) 	\\
 & (J2000) & (J2000) & & $[\times10^{44}$ erg s$^{-1}]$\\
\hline
Abell 968   & 10$^{\rm{h}}$21$^{\rm{m}}$09.5$^{\rm{s}}$ 	& +68$\degree$15$^{\prime}$53$^{\prime\prime}$ & 0.195 & 0.401 \\
Abell 981   & 10$^{\rm{h}}$24$^{\rm{m}}$24.8$^{\rm{s}}$ 	& +68$\degree$06$^{\prime}$47$^{\prime\prime}$ & 0.202 & 1.670 \\
Abell 998   & 10$^{\rm{h}}$26$^{\rm{m}}$17.0$^{\rm{s}}$ 	& +67$\degree$57$^{\prime}$44$^{\prime\prime}$ & 0.203 & 0.411 \\
Abell 1005 & 10$^{\rm{h}}$27$^{\rm{m}}$29.1$^{\rm{s}}$ 	& +68$\degree$13$^{\prime}$42$^{\prime\prime}$ & 0.200 & 0.268 \\
Abell 1006 & 10$^{\rm{h}}$27$^{\rm{m}}$37.2$^{\rm{s}}$ 	& +67$\degree$02$^{\prime}$41$^{\prime\prime}$ & 0.204 & 1.320 \\
\hline
\end{tabular}
References:
\begin{tablenotes}
	\item{Redshift, $z$: \citet{1990ApJ...365...66H}}
	\item{X-ray luminosity, $L_{\rm{x}}$: BAX database; \citet{2004A&A...424.1097S}}
\end{tablenotes}
\end{threeparttable}

\end{center}
\end{table}

\section{Observations \& Data Reduction}\label{sec:obs}
\subsection{Technical Summary}
The AMI telescope \citep{zwart2008} has recently undergone a correlator upgrade, and is now equipped with a wide-band digital back-end that possesses 1.2~MHz spectral resolution; for full details of the upgraded instrument, see \cite{hickish2017}. Here, we will briefly summarise the relevant technical information. The AMI-LA comprises eight 13-metre antennas with baselines in the range 18-110 metres, located near Cambridge, UK. AMI operates at a central frequency of 15.5~GHz, with an effective bandwidth of 5~GHz. This bandwidth is subdivided into two spectral windows centred at 14.25 and 16.75~GHz, each comprising $1024$ channels. The primary beam FWHM at 15.5~GHz is 5.5~arcminutes, and the typical resolution is of the order of $\sim30-40$ arcsec depending on the $uv$-coverage.

\subsection{Observing Details}
The SuperCLASS field was observed with the AMI-LA between 2016 July and 2017 May. The survey field was chosen to cover a $\sim0.9$ square degree area encompassing the Northern four clusters of the SuperCLASS super-cluster (Abell 968, 981, 998 and 1005; see Table \ref{tab:clusters} for details). Given the AMI-LA primary beam FWHM, the survey field was divided into twelve sub-fields, each comprising 20 close-packed pointings in a rectangular five-by-four grid, spaced at 0.5 FWHM at 15.5~GHz. The AMI-LA PB FWHM varies between around 6.3 arcmin and 5.2 arcmin across the observing band. As such, we should still retain close to uniform sensitivity at the highest frequencies. In total, approximately $300$ hours of data were taken.

\begin{figure}
	\begin{center}
		\includegraphics[width=0.45\textwidth]{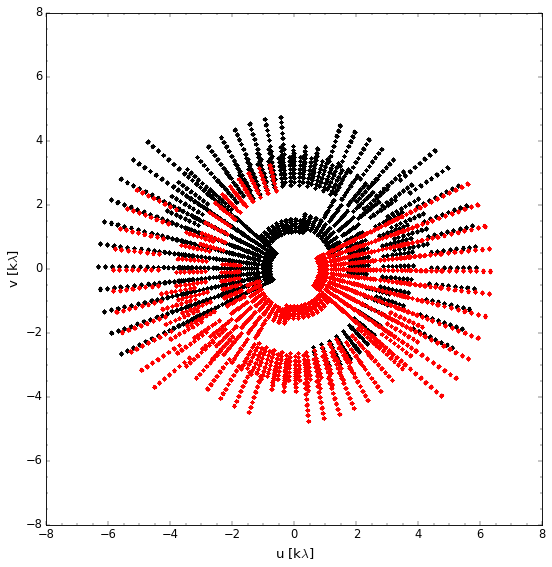} 
	\end{center}
\caption{Typical $uv$-coverage for a given AMI-LA pointing on the SuperCLASS field. Note that this is the final $uv$-coverage yielded by repeated observations over the course of several days. Different colours denote the mirrored $uv$ data from conjugate visibilities. \label{fig:uvcovg}}
\end{figure}

The AMI-LA typically observes two primary calibrators (3C~286 and 3C~48) for $\sim30$ minutes on a daily basis. Each sub-field was observed a number of times over the course of several days; on any given day, a typical observing run was between 4--8 hours in duration. We preferentially used 3C~286 to set the flux density scale; depending on availability and data quality, 3C~48 was substituted instead. Note that the primary factor affecting data quality is the weather. For our raster, we used a dwell time of $60$ seconds per pointing, revisiting all pointings multiple times in order to maximise the $uv$-coverage on each pointing. Three-minute scans of our phase calibrators (J1048+7143 or J0958+6533) were interleaved every 30 minutes. We present the final $uv$-coverage of a typical pointing in Figure \ref{fig:uvcovg}.

\subsection{Data Reduction}
The data were reduced using standard techniques for data reduction in CASA, with one exception: at these frequencies, a non-standard calibration table must be applied before standard calibration. This table (known as the `rain gauge') corrects for changes in the system temperature, which is monitored through use of a modulated noise source.

Calibration of AMI data is strongly affected by adverse weather. We found that several days' observations had to be completely discarded due to poor data quality, even after conservative flagging and rain gauge correction. In such cases, these fields were re-observed to minimize any losses in sensitivity across the survey area. Following calibration, each sub-field was imaged on a per-day basis as a quality control check. Our data were tied to the \cite{perleybutler2013} flux density scale.

All good data were then exported for imaging using the NRAO Astronomical Image Processing Software (\texttt{AIPS}) as CASA does not yet support manual definition of the primary beam. The concatenated calibrated data for each individual pointing were split and imaged (using \texttt{IMAGR}) separately, at a central frequency of 15.5~GHz.

We used a cell size of 4 arcseconds to adequately oversample the restoring beam; each pointing was mapped as a $512\times512$ pixel region. We employed natural weighting (\texttt{AIPS} \texttt{ROBUST} +5) to maximise our signal-to-noise. In order to mitigate clean bias, we restricted \texttt{IMAGR} to positive clean components above a $150~\umu$Jy threshold. 

Given the differences in $uv$-coverage, the restoring beam varies on a per-pointing basis -- the beam major axis varies between 36 and 46 arcsec; the minor axes are more consistent at 26 to 29 arcsec. We convolved all pointing maps to a final resolution of $50\times32$ arcsec in order to ensure a uniform point-source response across the field. 

\begin{figure*}
	\begin{center}
		\includegraphics[width=1.0\textwidth]{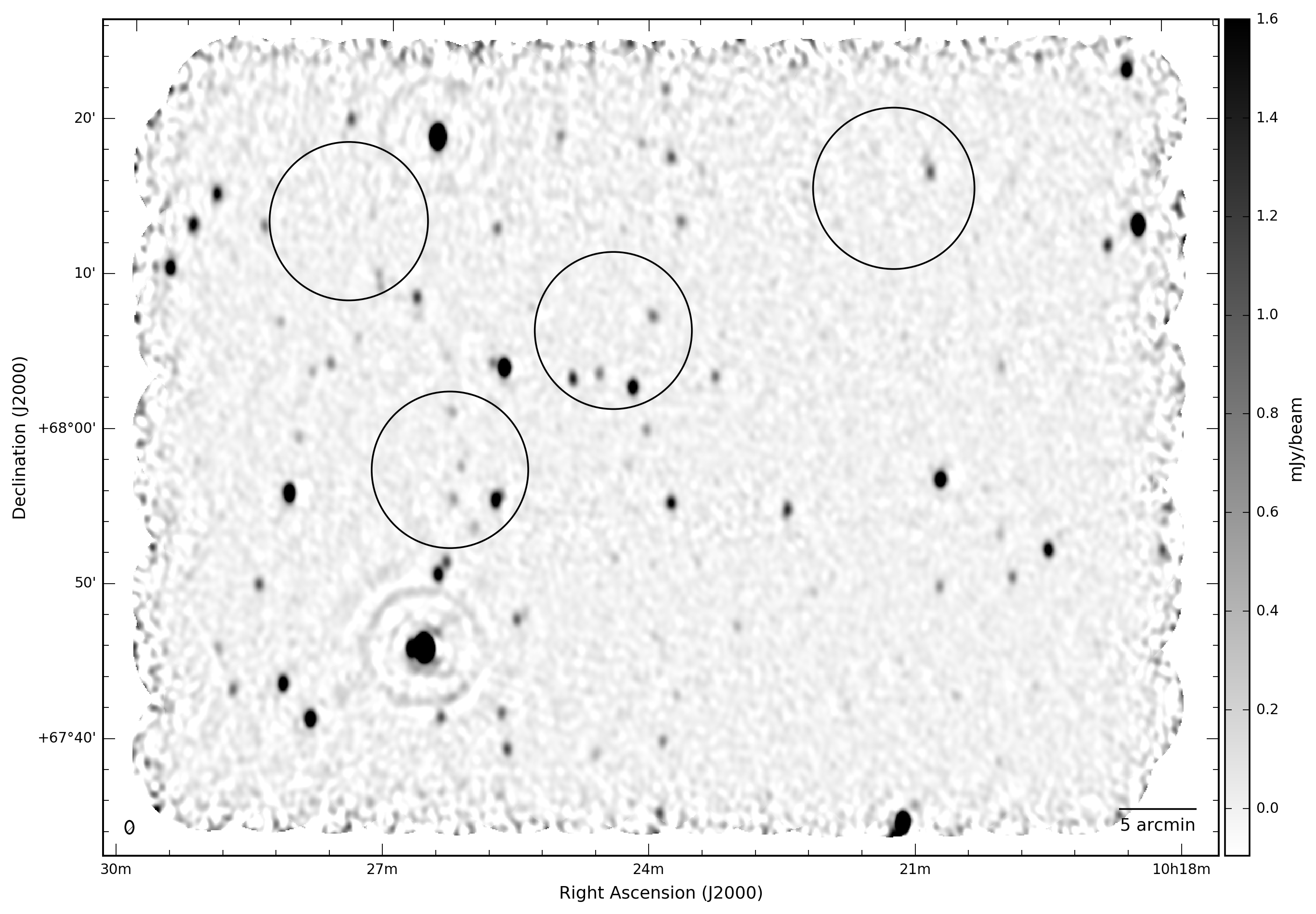}
	\end{center}
\caption{AMI-LA mosaic of the SuperCLASS field. The colour scale ranges from $-3\sigma$ to $50\sigma$, where $\sigma=32.0~\umu$Jy beam$^{-1}$ is the representative off-source image noise. The resolution is $50\times32$ arcsec, indicated by the ellipse in the lower-left corner. Circles of 1 Mpc radius are centred on Abell 968, Abell 981, Abell 998, Abell 1005 (right-to-left; see Table~\ref{tab:clusters}).}
\label{fig:surveyarea}
\end{figure*}

Subsequently, the noise level was measured on a per-pointing basis using \texttt{IMEAN}. We then used \texttt{FLATN} to combine the individual pointing maps, weighted according to the inverse square of the noise measured by \texttt{IMEAN}, above a primary beam cutoff of 10 per cent. We present our final mosaic of the survey area in Figure~\ref{fig:surveyarea}; a mosaic of the noise measured by \texttt{IMEAN} is presented in the left-hand panel of Figure~\ref{fig:rms}.

%
%As such, we acknowledge there will be a small loss of sensitivity; however, as one goal of this survey is to investigate the SZ signal from the super-cluster -- using the SA data observed simultaneously with these LA observations -- we require an accurate model of the compact radio source population for subtraction. We concluded that a consistent point-source response across the survey area was worth the slight loss of resolution.

\section{Results}\label{sec:res}
From Figure~\ref{fig:surveyarea}, the general image quality appears good. However, there are two bright sources in the field. Self-calibration of pointings in the vicinity of these sources has reduced the impact of sidelobes, although the dynamic range is limited to $\sim2\times10^3$. Our previous higher-resolution GMRT observations \citep{riseley2016} reveal the bright source visible in the South-East (SE) of Figure~\ref{fig:surveyarea} to be complex, exhibiting a dumbbell morphology, with a third compact source nearby. At the resolution of the AMI-LA, these three components blend into a compact double source. The bright source to the North-East (NE) of Figure~\ref{fig:surveyarea} is a single source both at the resolution of the AMI-LA and the GMRT.

Also visible in Figure~\ref{fig:surveyarea} are noise enhancements toward the edge of the mosaic - these are due to the rapid drop-off in the primary beam sensitivity, rather than anything astronomical. A small number of radio sources appear coincident with the four Abell clusters in this region; no redshift information is yet available for these sources, so we cannot determine whether these may be cluster member sources. All clusters in the SuperCLASS super-cluster are relaxed, exhibiting no evidence of merger events. As such, we do not expect the presence of any large-scale diffuse radio emission.

\subsection{Source Detection}
\begin{figure*}
	\begin{center}
		\includegraphics[width=\textwidth]{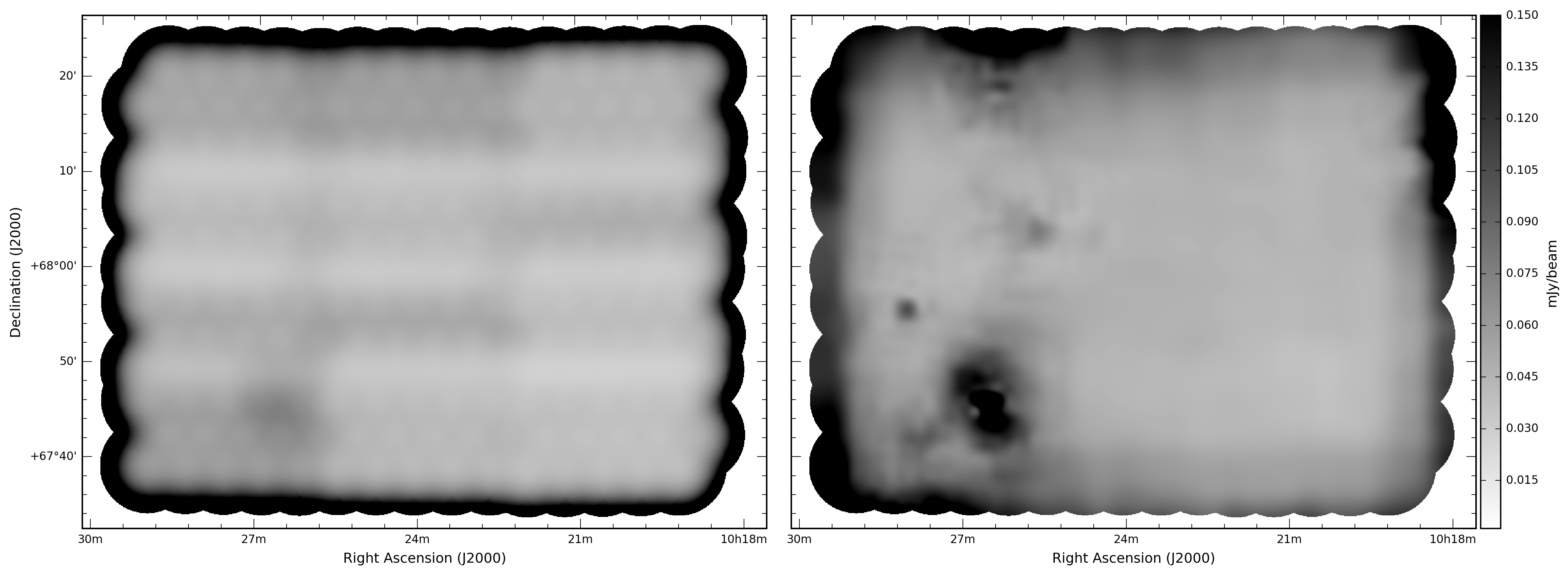} 
	\end{center}
\caption{Measured noise in the SuperCLASS field. \emph{Left panel:} noise map derived using \texttt{IMEAN} in \texttt{AIPS} for each pointing, and subsequently mosaicking using \texttt{FLATN}. \emph{Right panel:} 2D noise map derived by PyBDSM using a $({\rm{box, step}})$ size of $(150,30)$ pixels. The colour scale in each panel is identical, saturating at 0.15 mJy beam$^{-1}$.}
\label{fig:rms}
\end{figure*}

Sources were catalogued using the Python Blob Detection \& Source Measurement \citep[PyBDSM;][]{pybdsm} software\footnote{We note that with the most recent update, this has now been renamed PyBDSF -- the Python Blob Detection \& Source Finder. However, our catalogue was compiled before this update so we shall use the software as named at the time.}. PyBDSM uses a moving box to derive a map of the local rms (presented in Figure~\ref{fig:rms}), isolating islands above a user-defined threshold, and attempts to iteratively fit Gaussians to peaks above a given threshold. We characterized the rms using a $({\rm{box, step}})$ size of $(150,30)$ pixels for the majority of the image. We used PyBDSM's wavelet mode (\texttt{atrous\_do} = \texttt{True}) to decompose the residual image into wavelets on a small number of scales, in order to search for additional sources. No valid sources were found using the wavelet mode. 

We used an island (peak) threshold of $3\sigma$ $(5\sigma)$ to find sources, enabling the \texttt{adaptive\_threshold} mode to allow PyBDSM to better model the noise near bright sources. Near bright sources\footnote{Those with a signal-to-noise ratio in excess of 50.} a smaller $({\rm{box, step}})$ size of $(50,10)$ pixels was used. The 2D noise map derived by PyBDSM is presented in the right panel of Figure~\ref{fig:rms}. The map is generally similar to that measured by \texttt{IMEAN}, although areas of enhanced noise are clearly visible near the bright sources in the field.

PyBDSM initially catalogued 130 sources in our field; however, 42 of these lay in the field periphery where the signal-to-noise is significantly degraded. Inspection revealed that three of these corresponded to real radio sources (cases where a source was visible at the same location in our 325~MHz GMRT image) but were cut by the AMI-LA primary beam. The remainder could not be reliably identified with any source in our 325~MHz GMRT catalogue (hereafter referred to as the `SCG325 catalogue'; \citealt{riseley2016}) or the 1.4~GHz NVSS catalogue. As such, these sources were all removed from our catalogue.

\section{Catalogue Verification}\label{sec:an}

\subsection{Resolved Sources}\label{sec:unresolved}
\subsubsection{Method I: flux density ratio}
We can use the ratio of integrated to peak flux density to determine the spatial extent of a source \citep[e.g.][]{schinnerer2010,hales2014b}
\begin{equation}
	\frac{S_{\rm{int}}}{S_{\rm{peak}}} = \frac{\theta_{\rm{maj}}\theta_{\rm{min}}}{B_{\rm{maj}}B_{\rm{min}}}
\end{equation}
where $\theta_{\rm{maj}}$ and $\theta_{\rm{min}}$ are the $observed$ major and minor axes, and $B_{\rm{maj}}$ and $B_{\rm{min}}$ are the major and minor axes of the restoring beam. For a truly unresolved source, this ratio should equal unity. However, image noise may bias the fit, and the recovered peak and integrated flux densities of unresolved sources may differ.

To maintain consistency with our previous work at 325~MHz, we attempt to correct for this effect by defining a locus that envelops $99\%$ of sources with $S_{\rm{peak}}/S_{\rm{int}}>1$ as a function of detection significance. Assuming that a similar number of unresolved sources will be biased $high$ by this effect as are biased $low$ (yielding $S_{\rm{peak}}/S_{\rm{int}}>1$) we can mirror this locus above $S_{\rm{peak}}/S_{\rm{int}}=1$. 

\begin{figure}
	\begin{center}
		\includegraphics[width=0.45\textwidth]{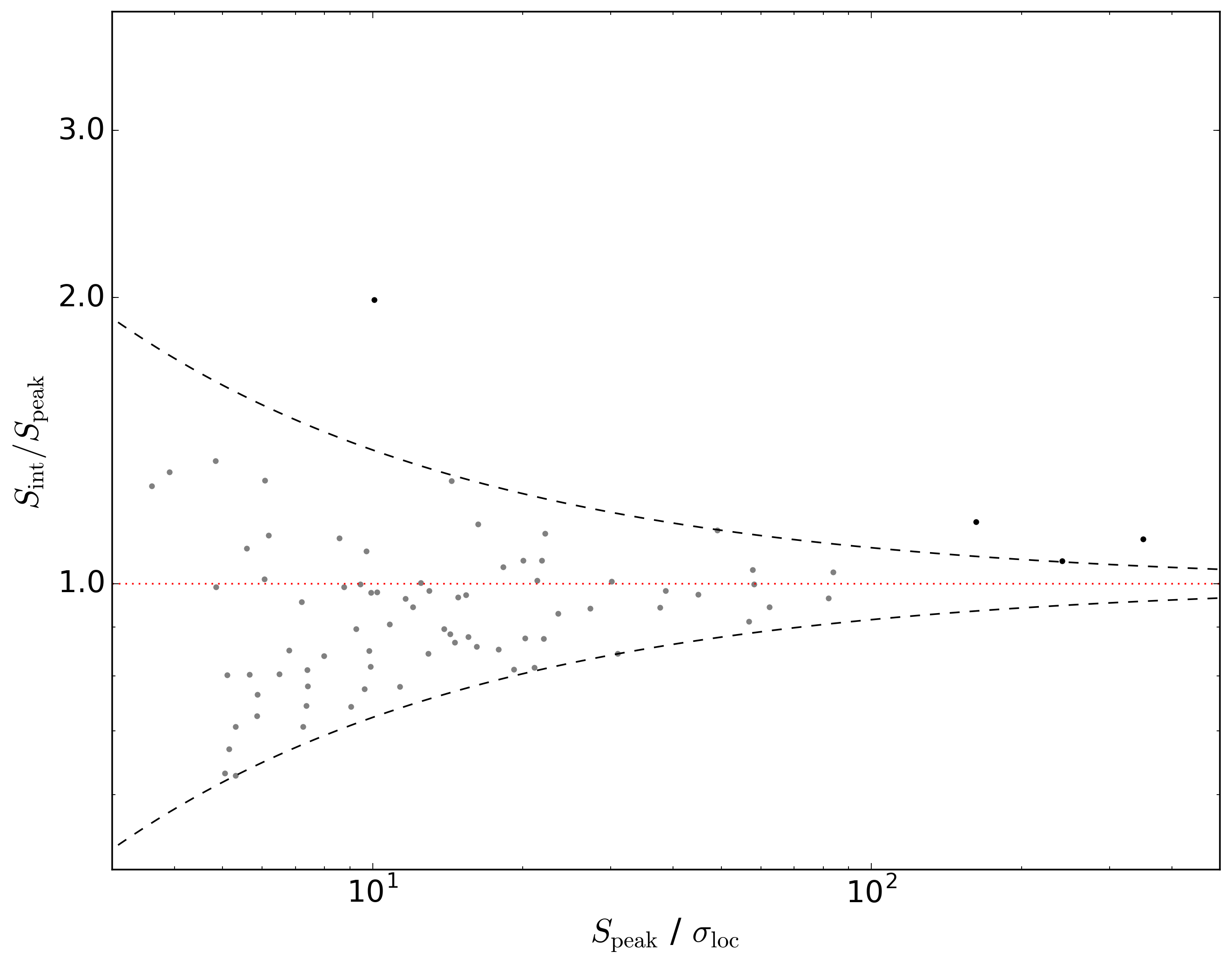} 
	\end{center}
\caption{Ratio of integrated to peak flux density as a function of detection significance. Red dotted line denoted a flux density ratio equal to unity. Dashed curves mark the locus defined by Equation~\ref{eq:locus}. Sources within this locus are defined as unresolved (grey points); sources above this locus are defined as resolved (black points).}
\label{fig:unresolved}
\end{figure}

Our locus is defined by 
\begin{equation}\label{eq:locus}
	\frac{S_{\rm{peak}}}{S_{\rm{int}}} = k^{(S_{\rm{peak}} / \sigma_{\rm{loc}})^{-c}}
\end{equation}
where $k=3.33$ and $c=0.57$ provide the best fit to our data. Sources that lie above this locus would be considered `resolved', with those that lie between these loci listed as unresolved. This fit is presented in Figure~\ref{fig:unresolved}. From Figure~\ref{fig:unresolved}, it is clear that (by this metric) the majority of sources in our catalogue are unresolved, as 75 are classified as unresolved, compared to only 5 resolved sources. It is also clear that the SNR is modest -- the highest detection significance is $S_{\rm{peak}}/\sigma_{\rm{loc}}\sim400$ (c.f. the SCG325 catalogue, where the highest detection significance was $\sim4\times10^3$; \citealt{riseley2016})

However, for two of these five `resolved' sources, PyBDSM reports zero size. For the remaining three, inspection reveals that two are badly fit - one is adversely affected by edge-of-field noise, and the other is clearly a faint double source (we discuss double sources later) that is modelled by PyBDSM as a single highly-elliptical Gaussian. The final `resolved' source is also clearly compact, from inspection, and is also compact in the SCG325 catalogue. We suggest that while this metric provided good constraining power at 325~MHz, the same cannot be said here -- likely because the SNR tends to be lower.

\subsubsection{Method II: critical deconvolved size}
Using simulations of synthetic point sources, \cite{franzen2011} showed that the relation $\beta \equiv \theta_{\rm{maj}} \rho^{1/2} / B_{\rm{maj}}$ could be used as an indicator of source extension. Here, $\rho$ is defined as $S_{\rm{peak}}/\sigma_{\rm{local}}$ -- i.e. the SNR. For the 10C survey, the critical size $(e_{\rm{crit}})$ that would discriminate between point-like and extended sources was
\begin{equation}
	e_{\rm{crit}} = \left\{
	\begin{array}{ll}
		3 B_{\rm{maj}} \rho^{1/2} & {\rm{if }}~3 B_{\rm{maj}} \rho^{1/2} > 25~{\rm{arcsec}} \\
		25~{\rm{arcsec}} & {\rm{otherwise}}
	\end{array}
	\right.
\end{equation}
where $B_{\rm{maj}}=50~{\rm{arcsec}}$ is the major axis of the restoring beam. Sources with $\theta_{\rm{maj}} > e_{\rm{crit}}$ were then defined as extended. \cite{whittam2016} use the same metric to separate resolved and unresolved sources.

By this metric, none of the sources in our catalogue are resolved -- all have deconvolved sizes that are up to an order of magnitude below $e_{\rm{crit}}$. As such, we will use the peak flux density in place of the integrated flux density for all sources in our catalogue.

\subsubsection{A note on double sources}
From Figure~\ref{fig:surveyarea}, there are six double sources visible. PyBDSM catalogues four of these as two separate compact sources, so using the individual peak flux densities in our catalogue should be appropriately representative. We present postage stamp images of all double sources in Figure \ref{fig:doubles}. However, two sources (the bright double source to the SE of the image and the faint double source located at about 1~Mpc from A1005) are merged by PyBDSM; as such, modelling these using simply the peak flux density would not be appropriate.

We have manually fitted these sources with \texttt{imfit} in CASA, using two components. For each source, the fits yield point-source components with flux densities consistent with that measured in the image plane, and noise-like residuals. We list the four components for each of these double sources separately in our catalogue. 

\subsection{Catalogue description}
Our final catalogue comprises 80 sources with flux densities in the range $193~\umu$Jy to $69.2~$mJy. Table~\ref{tab:src_cat} presents a sample of twenty sources from our 15.5~GHz AMI-LA catalogue. Here follows a brief description of catalogue columns:

Column (0): Source name, following the nomenclature J\emph{hhmmss+ddmmss}.

Columns (1) and (2): J2000 right ascension and declination in sexagesimal format.

Columns (3) and (4): J2000 right ascension and declination in decimal degrees.

Columns (5) and (6):  Flux density at 15.5~GHz, with the associated uncertainty.  

Columns (7) and (8): NVSS flux density at 1.4~GHz, with the associated uncertainty. These measurements have been adjusted to the \cite{perleybutler2013} scale using a correction factor 1.006 (the mean ratio of flux densities between the \cite{baars1977} and \cite{perleybutler2013} scales for 3C~286 and 3C~48). 

Columns (9) and (10): Integrated flux density at 325~MHz along with the associated uncertainty \citep[from][]{riseley2016}. These have had a factor 1.071 applied to convert from the \cite{scaifeheald2012} scale to the \cite{perleybutler2013} scale.

Columns (11) and (12): Spectral index between 325~MHz and 1.4~GHz, with the associated uncertainty. This is only derived for sources common to all three catalogues. 

Columns (13) and (14): Spectral index between 1.4~GHz and 15.5~GHz, with the associated uncertainty. This is only derived for sources common to all three catalogues. 

Column (15): Comment on whether multiple matches exist between catalogues.

\subsection{Completeness}
\subsubsection{Simulations \& visibility area}
There are a number of ways of assessing the completeness of our catalogue. For our survey -- as with our previous work at 325~MHz -- we have established the completeness empirically using a simulated compact source population. We established 20 flux density bins log-spaced between 0.2~mJy (consistent with the minimum flux density of source in our catalogue) and 2.5~mJy.

We also randomly selected the positions of 250 sources -- which were held constant for each flux density -- and added our simulated sources to the residual map produced by PyBDSM using \texttt{IMMOD} in \texttt{AIPS}. We restricted the placement of each source slightly so that i) sources would reside entirely within the residual image and ii) no source would be placed within 2~arcmin of another source. The relatively small field area, combined with these restraints, made it necessary to generate ten different position catalogues of 25 sources for each flux density value.

\begin{figure}
	\begin{center}
		\includegraphics[width=0.5\textwidth]{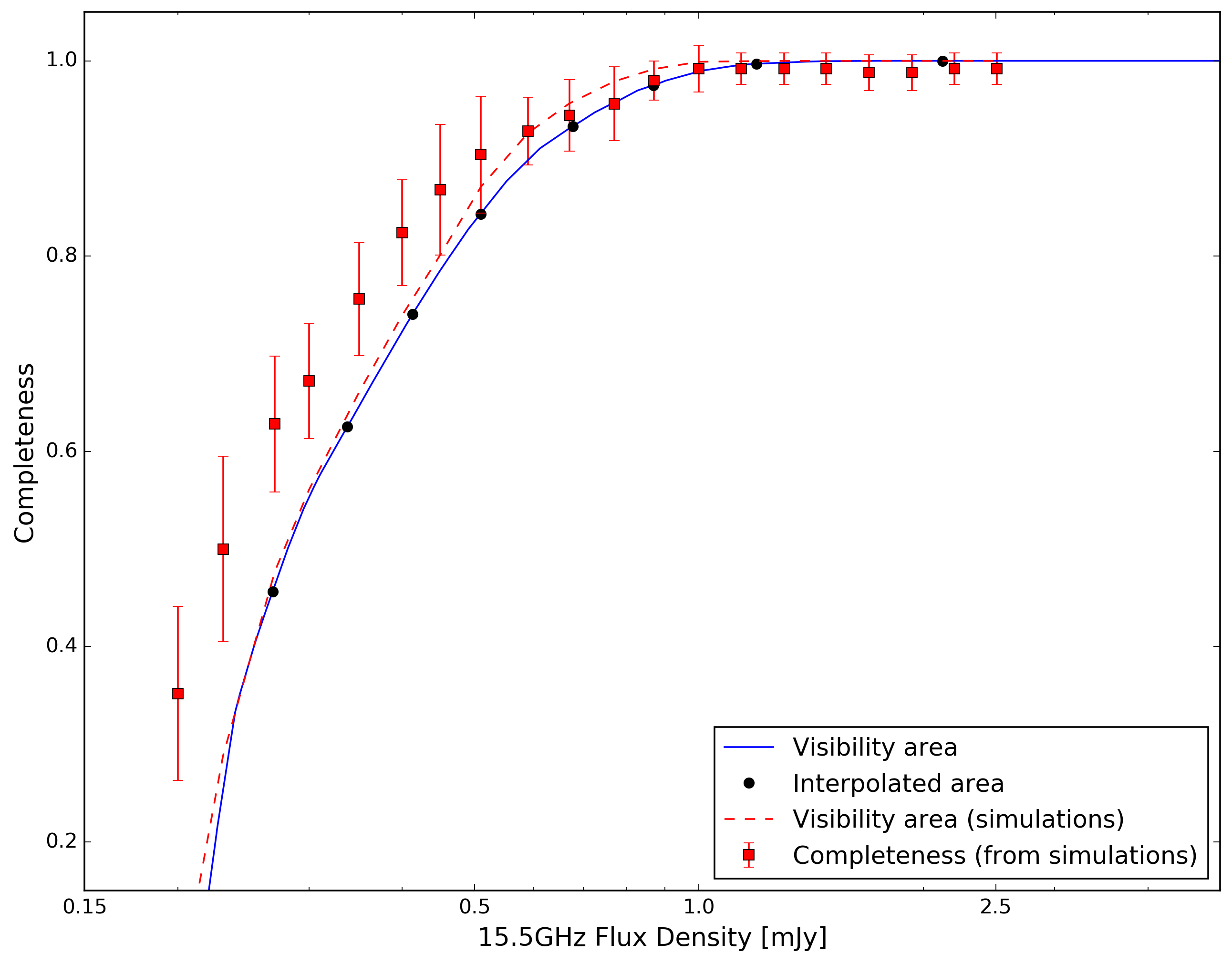} \\
	\end{center}
\caption{Result of completeness simulations for our survey. Red points denote the fraction of sources recovered as a function of flux density; the error bars denote the standard deviation in the recovered fraction. We also plot the visibility area as a function of flux density, based on the rms noise profile in Figure~\ref{fig:rms} (in blue) assuming Gaussian statistics. The dashed red line denotes the visibility area of the rms maps derived when evaluating the completeness simulations. Black points are the visibility area interpolated to the bin central flux densities in Table~\ref{tab:src_counts} (see \S\ref{sec:src}).}
\label{fig:visarea}
\end{figure}

Once these catalogues were produced, sources were catalogued by PyBDSM, using identical settings to our real AMI-LA image. The recovered sources were then cross-matched with the known positions, and deemed to match if a source was recovered within 30~arcsec of its known position. We present the fraction of sources recovered in Figure~\ref{fig:visarea}, where the error bars denote the standard deviation in the fraction of sources detected.

Additionally presented in Figure~\ref{fig:visarea} is the visibility area -- the fraction of the image over which a source of a given flux density $S_i$ should be detectable (i.e. $S_i \geq 5\sigma_{\rm{local}}$). This is derived using the rms map presented in the right panel of Figure~\ref{fig:rms}; from the visibility area, we estimate that we achieve 95\% completeness at a limiting flux density of 0.76~mJy. We also use the visibility area in \S\ref{sec:src} to derive the differential source count.

In Figure~\ref{fig:visarea} we also present the visibility area of the rms maps derived by PyBDSM during this process. As can be seen in Figure~\ref{fig:visarea}, the two visibility area measurements are very similar. Figure~\ref{fig:visarea} suggests that the completeness is slightly higher than predicted by the visibility area toward the limit of our survey; at flux densities of a few mJy, our simulations are consistent with the visibility area calculation. This same effect was also seen in completeness simulations for the 10C survey \citep[e.g.][]{davies2011}.

A natural explanation for this effect might be source confusion, as two blended sources below the nominal flux density limit could appear sufficiently bright to be detectable. However, this cannot be the case here, as we ensured a minimum separation of 2~arcmin between simulated sources. 

\begin{landscape}
\begin{table}
\caption{Excerpt from our 15.5~GHZ AMI-LA catalogue. Column (0): Source name, following the nomenclature J\emph{hhmmss+ddmmss}. Columns (1) and (2): Right ascension and declination in sexagesimal format, J2000 reference. Columns (3) and (4): J2000 right ascension and declination in degrees. Columns (5) and (6): Flux density at 15.5~GHz, with the associated uncertainty. Columns (7) and (8): Flux density at 1.4~GHz, from the NVSS, with the associated uncertainty. Columns (9) and (10): Integrated flux density at 325~MHz along with the associated uncertainty. Columns (11) and (12): Spectral index between 15.5~GHz and 1.4~GHz, with the associated uncertainty. Columns (13) and (14): Spectral index between 325~MHz and 1.4~GHz, with the associated uncertainty. Column (15) marks whether sources have multiple and/or complex matches at 325~MHz. Columns (9) through (12) are only filled for sources present in all three catalogues. Flux density measurements in columns (5) through (8) have been adjusted to the \protect\cite{perleybutler2013} flux density scale. }
\label{tab:src_cat}

\begin{tabular}{ccc|cc|rr|rr|rr|rr|rr|c}
\hline
(0) & (1) & (2) & (3) & (4) & (5) & (6) & (7) & (8) & (9) & (10) & (11) & (12) & (13) & (14) & (15) \\
Source name & RA (J2000) & DEC (J2000) & RA (J2000) & DEC (J2000) & \multicolumn{2}{c|}{$S_{15.5~{\rm{GHz}}}$} & \multicolumn{2}{c|}{$S_{1.4~{\rm{GHz}}}$} & \multicolumn{2}{c|}{$S_{325~{\rm{MHz}}}$} & \multicolumn{2}{c|}{$\alpha_{\rm{high}}$} & \multicolumn{2}{c|}{$\alpha_{\rm{low}}$} & Comments \\
& \emph{hh mm ss.ss} & \emph{dd mm ss.s} & \emph{degrees} & \emph{degrees} & \multicolumn{2}{c|}{$[$mJy$]$}  & \multicolumn{2}{c|}{$[$mJy$]$} &  \multicolumn{2}{c|}{$[$mJy$]$} & & & & \\
\hline
J102932+681028 & 10 29 32.36 & 68 10 28.29 & 157.38482 & 68.17453 & 2.88 & 0.20 & 50.50 & 1.61 & 144.58 & 10.22 & $-1.19$ & 0.03 & $-0.72$ & 0.05 & \\
J102917+681316 & 10 29 17.09 & 68 13 16.53 & 157.32121 & 68.22126 & 1.88 & 0.15 & 19.32 & 1.01 & 61.02 & 4.32 & $-0.97$ & 0.04 & $-0.79$ & 0.06 & \\
J102900+681517 & 10 29 00.98 & 68 15 17.92 & 157.25407 & 68.25498 & 1.89 & 0.13 & 4.23 & 0.40 & 3.01 & 0.24 & $-0.34$ & 0.05 & 0.23 & 0.08 & \\ 
J102853+674559 & 10 28 53.66 & 67 45 59.96 & 157.22359 & 67.76666 & 0.53 & 0.07 & - & - & 0.39 & 0.09 & - & - & - & - \\
J102843+674321 & 10 28 43.10 & 67 43 21.11 & 157.17958 & 67.72253 & 0.88 & 0.08 & - & - & 6.67 & 0.48 & - & - & - & - \\
J102826+675008 & 10 28 26.73 & 67 50 08.53 & 157.11138 & 67.83570 & 1.07 & 0.07 & 3.32 & 0.50 & 5.73 & 0.45 & $-0.47$ & 0.07 & -0.37 & 0.11 & \\
J102827+681316 & 10 28 27.28 & 68 13 16.46 & 157.11368 & 68.22124 & 0.79 & 0.06 & - & - & 2.65 & 0.20 &  - & - & - & - \\
J102815+680706 & 10 28 15.39 & 68 07 06.24 & 157.06414 & 68.11840 & 0.45 & 0.05 & 3.02 & 0.50 & 6.62 & 0.50 & $-0.79$ & 0.08 & $-0.54$ & 0.12 & \\
J102809+674346 & 10 28 09.13 & 67 43 46.89 & 157.03803 & 67.72969 & 2.98 & 0.17 & 16.40 & 0.60 & 26.57 & 1.89 & $-0.71$ & 0.03 & $-0.33$ & 0.06 & \\
J102807+675604 & 10 28 07.17 & 67 56 04.46 & 157.02988 & 67.93457 & 5.01 & 0.26 & 2.11 & 0.40 & 0.58 & 0.10 & 0.36 & 0.08 & 0.89 & 0.17 & \\
J102801+675941 & 10 28 01.13 & 67 59 41.69 & 157.00470 & 67.99491 & 0.46 & 0.05 & - & - & 3.44 & 0.25 & - & - & - & - \\
J102750+674132 & 10 27 50.23 & 67 41 32.29 & 156.95930 & 67.69230 & 4.20 & 0.23 & 52.21 & 1.61 & 170.59 & 12.08 & $-1.05$ & 0.03 & $-0.81$ & 0.05 & \\
J102752+680356 & 10 27 52.58 & 68 03 56.93 & 156.96907 & 68.06581 & 0.45 & 0.05 & - & - & 2.43 & 0.19 & - & - & - & - \\ 
J102740+680429 & 10 27 40.07 & 68 04 29.36 & 156.91696 & 68.07482 & 0.73 & 0.06 & 2.82 & 0.40 & 5.56 & 0.41 & $-0.56$ & 0.07 & $-0.47$ & 0.11 & \\
J102728+682015 & 10 27 28.20 & 68 20 15.05 & 156.86750 & 68.33751 & 1.08 & 0.10 & - & - & 0.53 & 0.11 & - & - & - & - \\
J102721+680609 & 10 27 21.34 & 68 06 09.34 & 156.83890 & 68.10259 & 0.33 & 0.04 & - & - & 0.45 & 0.09 & - & - & - & - \\
J102640+680944 & 10 26 40.96 & 68 08 49.44 & 156.67065 & 68.14707 & 1.23 & 0.08 & 2.41 & 0.40 & 2.64 & 0.20 & $-0.28$ & 0.07 & $-0.06$ & 0.13 & \\
J102640+680737 & 10 26 40.12 & 68 07 37.21 & 156.66715 & 68.12700 & 0.25 & 0.06 & - & - & - & - & - & - & - & - \\
J102627+681911 & 10 26 27.74 & 68 19 11.40 & 156.61557 & 68.31983 & 23.20 & 1.16 & 206.83 & 6.24 & 660.90 & 46.73 & $-0.91$ & 0.02 & $-0.80$ & 0.05 & \\
J102618+675145 & 10 26 18.83 & 67 51 45.78 & 156.57847 & 67.86272 & 1.14 & 0.11 & 2.82 & 0.50 & 3.45 & 0.19 & - & - & - & - & $\ast$ \dag \ddag  \\
J102624+675057 & 10 26 24.38 & 67 50 57.96 & 156.60157 & 67.84943 & 2.50 & 0.15 & 2.82 & 0.50 & 3.90 & 0.39 & - & - & - & - & $\ast$ \dag \ddag  \\
J102621+674144 & 10 26 21.45 & 67 41 44.27 & 156.58936 & 67.69563 & 1.07 & 0.12 & 12.88 & 0.60 & 54.49 & 3.86 & $-1.04$ & 0.05 & $-0.99$ & 0.06 & \\
\hline
\end{tabular}
\begin{tablenotes}
	\item{\dag: Denotes multiple AMI-LA sources with a single NVSS match.}
	\item{\ddag: Denotes multiple/complex matches with SCG325 sources.}
	\item{$\ast$: NVSS cross-match may be spurious. Source omitted from colour-colour plot }
\end{tablenotes}
\end{table}
\end{landscape}

\subsubsection{Expected number of sources}
\cite{Whittam2017} present a correction for the S$^3$ source count that accounts for the discrepancy between the observed 10C faint source count and those predicted by S$^3$. The correction includes an enhanced core fraction in the flux density distribution for low-luminosity FRI sources. Integrating the corrected source count between our 95\% completeness limit and the flux density limit of our survey (69.2~mJy) suggests we should recover 42 sources, given our survey area (0.83 square degrees). 

Our catalogue contains 40 sources in this flux density range, which suggests there is no evidence for an enhanced source density. This is slightly unexpected, as our survey covers a galaxy super-cluster; the lack of overdensity would suggest that we are still dominated by the field population rather than that of the super-cluster. We further consider the distribution of our sources on the sky in \S\ref{sec:clustering}, where we perform a 2D clustering analysis.

\subsection{Flux Density Verification}
During the 10C survey, the AMI-LA was not sufficiently stable for the flux scale to be set using daily observations of known calibrators \citep[e.g.][]{franzen2011}. However, with the recent correlator upgrade, both the LA and SA flux density scales are tied solely to daily observations of 3C~286 and 3C~48. Our flux density scale was tied to 3C~286 where possible, or 3C~48 where data for 3C~286 were unavailable/did not yield good calibration solutions. All sources were tied to the \cite{perleybutler2013} flux density scale. 

For the 10C survey \citep[e.g.][]{davies2011} and the Ultra-deep 10C extension \citep{whittam2016} the signal-to-noise ratio (SNR) for recovered sources was too poor to successfully self-calibrate the survey data. Instead, they apply a bulk correction factor 1.082 to the catalogues based on the difference between the flux densities of bright sources recovered from self-calibrated, pointed observations and those recovered from the survey raster observations \citep[see][for details]{davies2011}. \cite{whittam2016} apply the same correction factor.

With the new digital back-end, however, the typical AMI-LA dynamic range has improved by approximately an order of magnitude \citep{hickish2017}. While the majority of sources in the field do not possess sufficient flux density to derive good self-calibration solutions, we have successfully performed phase-only self-calibration for the pointings that are closest to the brightest two sources in our field. This has significantly reduced the effect of sidelobes in the immediate area, although the dynamic range is still limited to $\sim2\times10^3$. We note that we measure consistent flux densities between our initial raster maps and the self-calibrated images.

\subsubsection{Verifying the calibration using the phase calibrator}
Interleaved calibrator sources at the frequencies considered here often exhibit a significant degree of variability. Given that our observations were conducted over several months, we can verify our calibration by comparing the flux densities measured by the AMI-LA with those measured by other observatories.

Two phase calibrators were observed during the course of our observing campaign -- J0958+6533 (hereafter J0958) and J1048+7143. The latter was only observed over the course of a few days in 2016 July; from 2016 August onwards, J0958 was used. As such, we will focus solely on J0958 here.

Since 2008, J0958 has been routinely observed as part of the OVRO 40-m monitoring campaign at 15~GHz \citep{Richards2011}. We have retrieved the flux density measurements for J0958\footnote{From \url{http://www.astro.caltech.edu/ovroblazars}} and these are plotted in Figure~\ref{fig:phasecalflux}, as well as our measurements with the AMI-LA. These are plotted in Figure~\ref{fig:phasecalflux} for the [MJD - 54466] range where our observations took place\footnote{The uncertainties on the AMI-LA flux density measurements are the quadrature sum of the measurement uncertainty plus five per cent of the flux density.}.

\subsubsection{Quantifying the uncertainty}
\begin{figure}
	\begin{center}
		\includegraphics[width=0.49\textwidth]{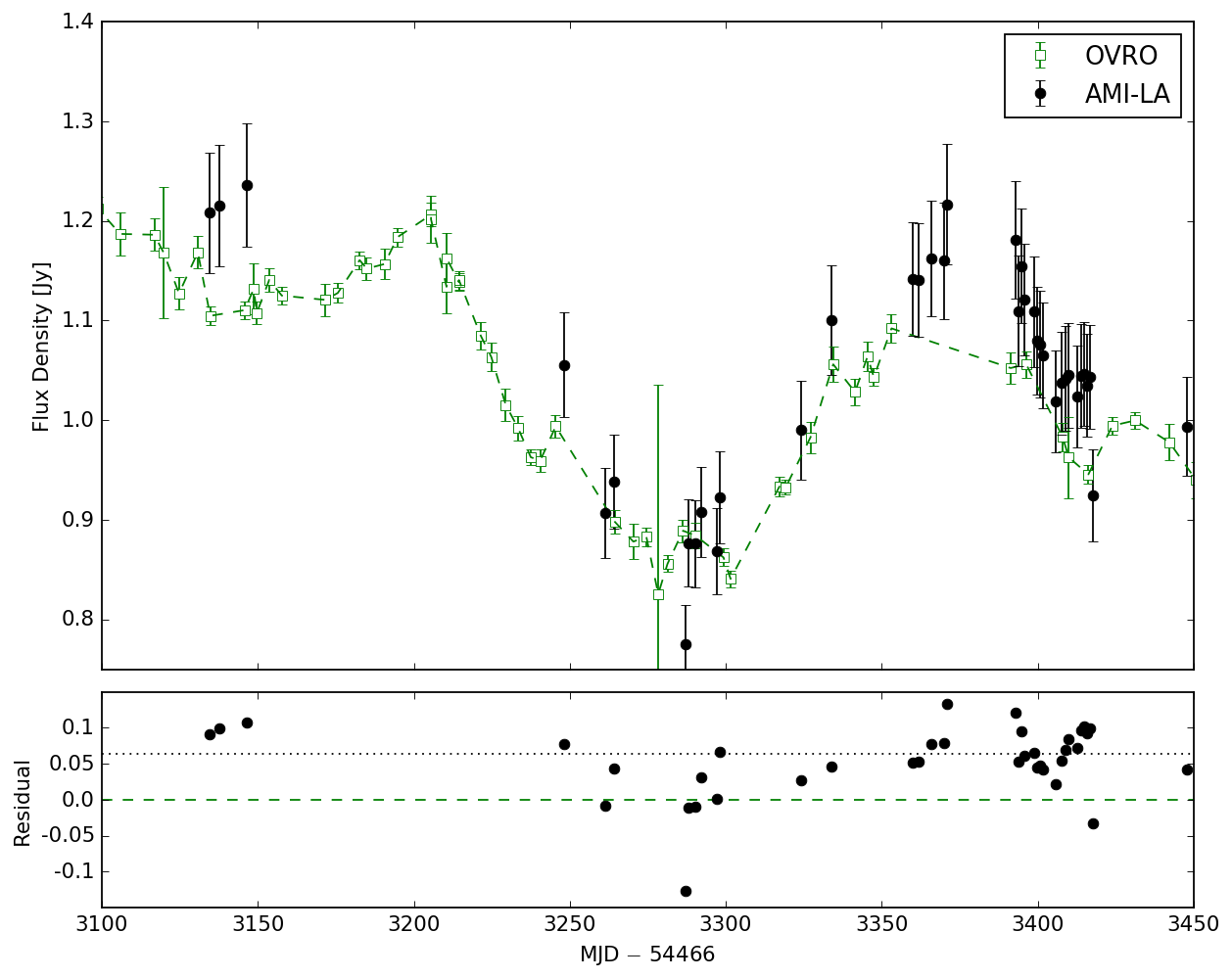} 
	\end{center}
\caption{Flux density measurements for J0958+6533 from the OVRO monitoring program \citep[green points;][]{Richards2011} and the AMI-LA (black points). Dashed green curve denotes spline fit to the OVRO measurements, used to derive the residuals (bottom panel). The mean absolute residual is indicated by the dotted line in the lower panel; dashed line denotes the OVRO spline fit (used as the reference).}
\label{fig:phasecalflux}
\end{figure}

In order to quantify the uncertainty on our flux density measurements, we have derived a spline fit to the OVRO flux density measurements for J0958. This is also presented in Figure~\ref{fig:phasecalflux}. In the lower panel, we plot the residual of our AMI-LA flux density measurements compared to the spline fit, for overlapping observation dates. From Figure~\ref{fig:phasecalflux} it is clear that J0958 has exhibited significant variability over the course of the monitoring period. We note that no OVRO observations took place during the [MJD-54466] period 3350 -- 3390; the trend in AMI-LA flux density measurements in this period suggest that a flare took place.

It is also apparent that the AMI-LA flux density measurements exhibit a bulk offset compared to the OVRO values. However, the AMI-LA collects only a single polarization (measuring I+Q) whereas OVRO collects dual-polarization data. This has been accounted for during calibration using the in-house version of the CASA task \texttt{setjy}, which corrects for the $Q/I$ fraction of 3C~286 and 3C~48 \citep[using the measurements of][]{PerleyButler2013B}. From \cite{Jackson2007} J0958 has a polarization fraction $Q/I = 7.2$ per cent at 8~GHz, which appears consistent with the bulk offset between the AMI-LA and OVRO measurements (the mean absolute residual is $6.4$ per cent). No polarization information is available for J0958 at 15.5~GHz, however, nor are its polarization properties routinely monitored.

From Figure~\ref{fig:phasecalflux}, the maximum absolute residual is of the order of 13 per cent; the standard deviation is $3.4$ per cent. As such, we will conservatively assume a five per cent calibration uncertainty on all flux density measurements in our catalogue. We therefore quote the uncertainty as five per cent of the flux density measurement plus the fitting uncertainty, added in quadrature. Given the consistency between the AMI-LA/OVRO offset and the \citeauthor{Jackson2007} polarisation measurement, we assume no overall systematic calibration offset.

\subsection{Ancillary data}
In the radio regime, this field has also been covered by the NRAO VLA Sky Survey \citep[NVSS;][]{condon1998} and 325~MHz SuperCLASS GMRT survey \citep[SCG325;][]{riseley2016}. In order to verify our catalogue as well as perform an initial investigation of the spectral index distribution across a factor 50 in frequency, we have cross-matched these catalogues. 

Given that our AMI-LA catalogue and the NVSS possess similar resolution (50~arcsec compared to 45~arcsec) we have used a cross-match radius of 50~arcsec. In our survey region, the NVSS catalogue contains approximately 65 sources above a limiting flux density of 2.1~mJy. We find a total of 43 matches between the NVSS and our AMI-LA catalogue, of which 41 are unique matches. There are three cases where a single NVSS source matches two AMI-LA sources.

Given the improved sensitivity of the SCG325 catalogue (nominally $34~\umu$Jy beam$^{-1}$ in this region), we would expect a greater number of matches with our AMI-LA catalogue. However, given the factor $\approx40$ difference in frequency (which means steep-spectrum sources undetected at 15.5~GHz may be in the SCG325 catalogue) and resolution (13~arcsec compared to 50~arcsec) we have used a smaller cross-match radius of 20~arcsec to reduce the number of potentially unrelated sources that match by chance. We find 74 AMI-LA sources which have counterparts in the SCG325 catalogue; of these, eight AMI-LA sources have two matches in the SCG325 catalogue. Additionally, there are six sources detected by the AMI-LA that have no counterparts at 325~MHz.

\subsubsection{Astrometry: comparison with the NVSS}
\begin{figure}
	\begin{center}
		\includegraphics[width=0.47\textwidth]{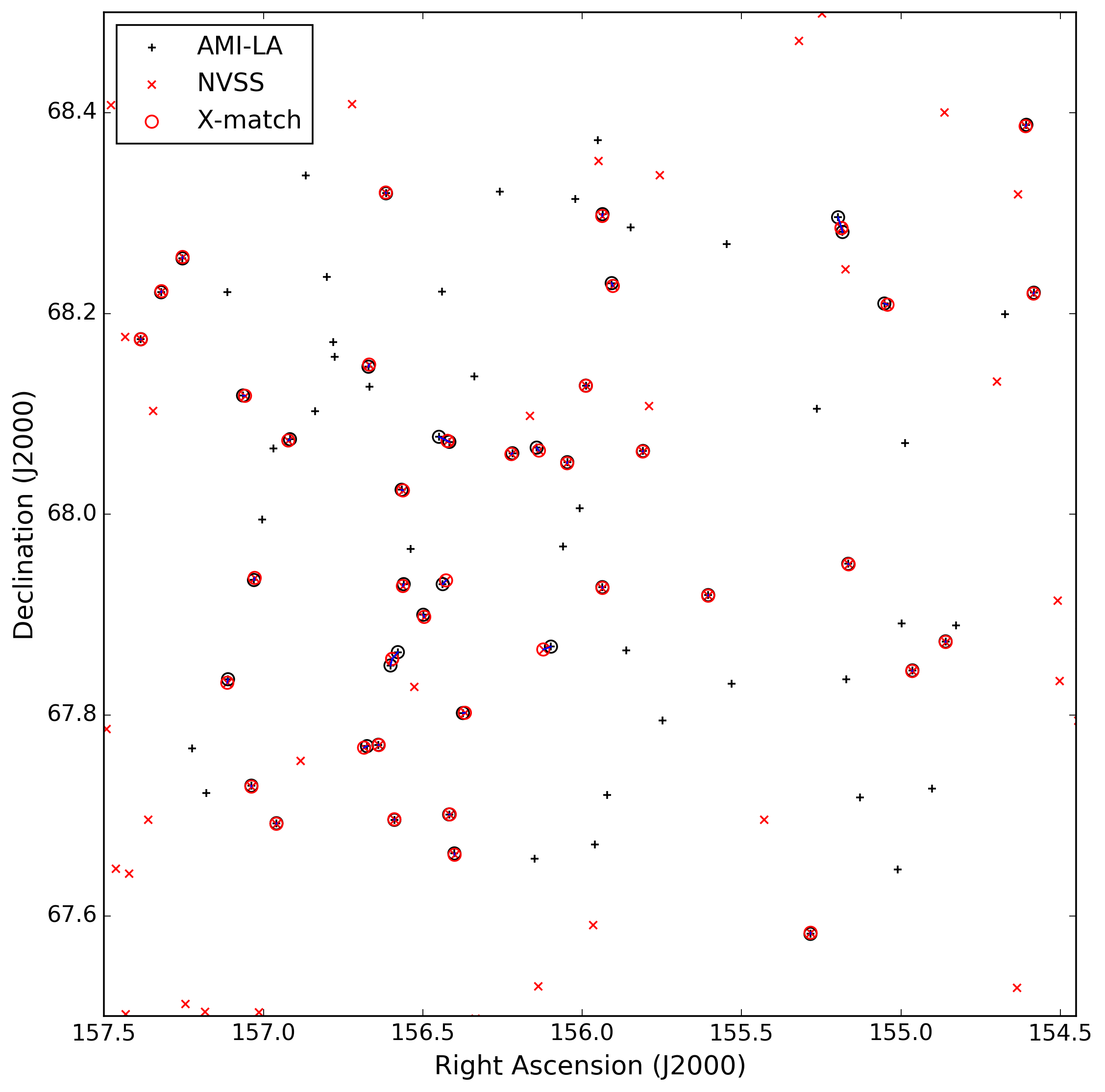} \\
		\includegraphics[width=0.47\textwidth]{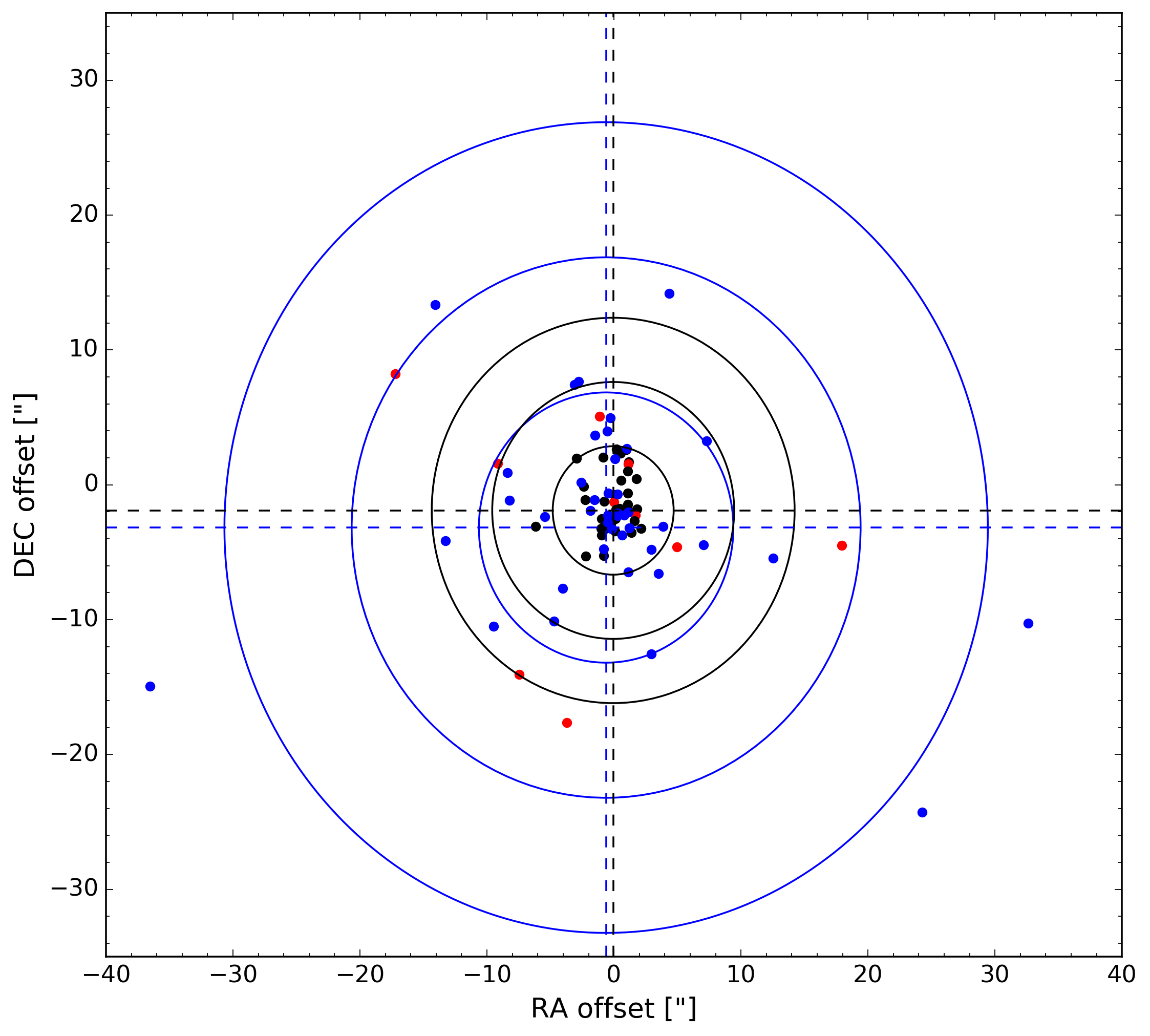} 
	\end{center}
\caption{\emph{Top:} Sky positions of sources in our AMI-LA catalogue (black `$+$') and the NVSS (red `$\times$'). Cross-matched sources are identified with red circles. \emph{Bottom:} positional offset of AMI-LA sources from the NVSS / SCG325 reference positions in blue / black. Concentric circles denote the $1,2,3\sigma$ offsets. Red points denote multiple SCG325 sources that match a single AMI-LA source.}
\label{fig:positions}
\end{figure}

The NVSS catalogue is known to have position accuracies better than 1 arcsec for sources brighter than 15 mJy, and better than 7 arcsec for fainter sources. We plot the positions of sources in our AMI-LA catalogue, as well as the NVSS catalogue and the cross matches, in the upper panel of Figure~\ref{fig:positions}. 

From Figure~\ref{fig:positions}, the majority of sources in our catalogue have counterparts in the NVSS. However, it is also clear that there are a significant number of AMI-LA sources that do not have NVSS counterparts. This may suggest a substantial population of sources with flat spectra -- this will be investigated later in \S\ref{sec:spix} -- although our AMI observations are a factor $\approx9$ deeper than the NVSS (assuming a spectral index $\alpha=-0.8$). 

\subsubsection{Astrometry: Discussion}
We also plot the astrometric offset between these catalogues in the lower panel of Figure~\ref{fig:positions}. The offsets for unique matches are shown in black; for multiple SCG325 sources that match a single AMI-LA source, the offsets are displayed in red. The astrometric offsets wrt. the NVSS positions are indicated by the blue points.

The mean offset between the fitted AMI-LA positions and the NVSS reference position is $\Delta({\rm{RA}})=-0.62$~arcsec and $\Delta({\rm{dec}})=-3.17$~arcsec. It can be seen in the lower panel of Figure~\ref{fig:positions} that the vast majority of sources exhibit very little offset. Three sources exhibit an offset in excess of 20~arcsec. However, from inspection, two of these cases occur where two AMI-LA sources match a single NVSS source. The final case originates from an AMI-LA cross-matching with a very faint NVSS source (peak flux density $\sim1.9~$mJy beam$^{-1}$) which may be spurious. 

Likewise, compared to the SCG325 catalogue, the mean offset is $-0.06$ arcsec in RA and $-1.90$ arcsec in dec. All sources which exhibit an offset outside the $3\sigma$ contour are the result of multiple cross-identifications; some of these associations may be spurious, given the factor 3.8 difference in resolution between instruments.

Given our resolution $(50\times32~{\rm{arcsec}})$ and pixel size $(4~{\rm{arcsec}})$ the source positions are generally consistent within $\sim2-3$ pixels of the NVSS position. Additionally, the mean astrometric offsets in both RA and dec are smaller than the mean uncertainties on the fitted positions in our AMI-LA catalogue (measured by PyBDSM). As such, we apply no correction to the position of sources in our catalogue.

\subsection{Spectral Index Distribution}\label{sec:spix}

\subsubsection{Matching with the NVSS}
\begin{figure}
	\begin{center}
		\includegraphics[width=0.47\textwidth]{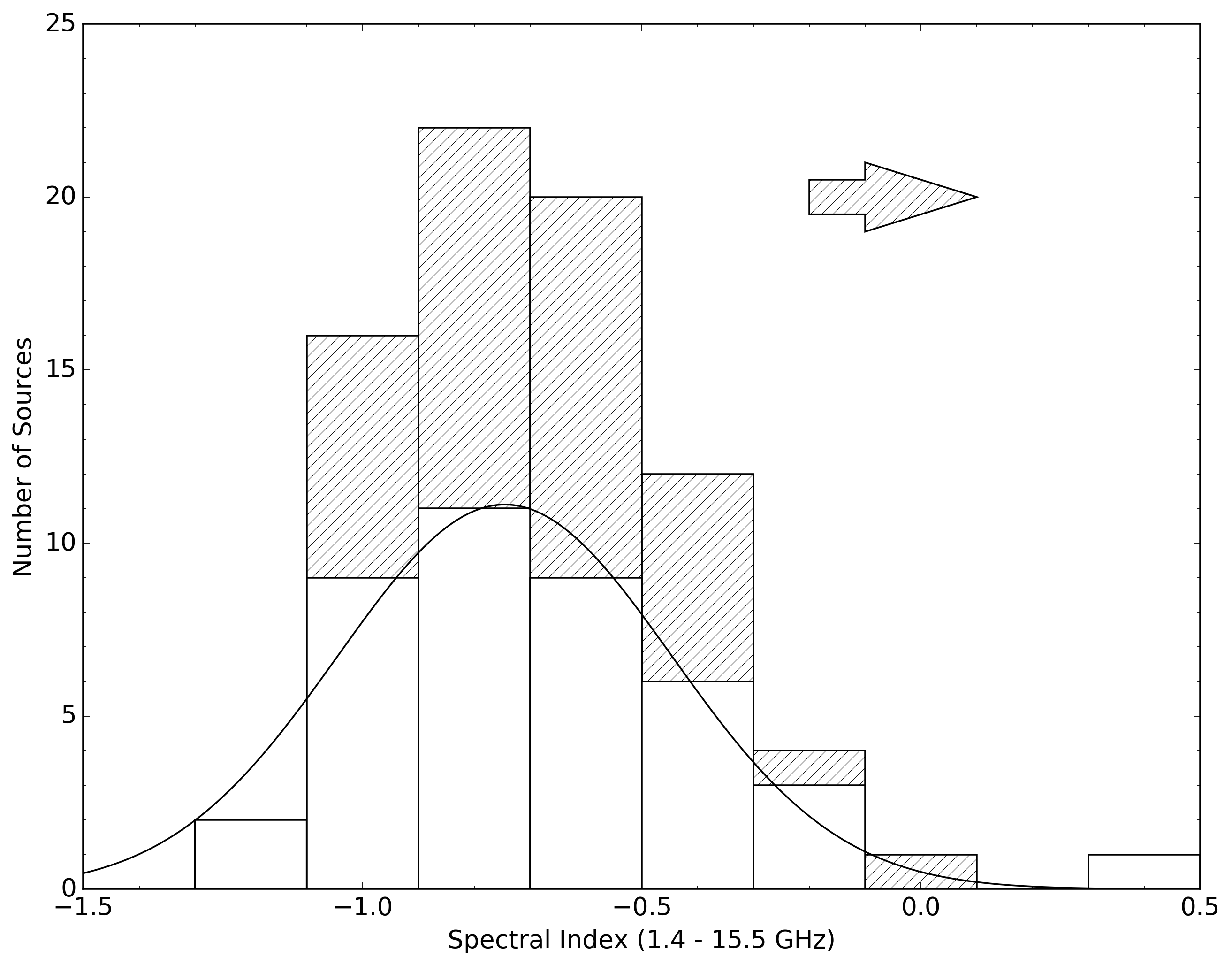} \\
	\end{center}
\caption{Spectral index distribution between 1.4 and 15.5~GHz for sources detected in our AMI-LA mosaic which have counterparts in the NVSS (open histogram) and lower-limit spectra for AMI-LA sources without NVSS counterparts (shaded histogram). The curve denotes the Gaussian fit to the measured spectral index distribution, centred on $\alpha_{\rm{high}} = -0.75\pm0.30$.}
\label{fig:spixdist}
\end{figure}

For sources in our 15.5~GHz catalogue which have counterparts in the NVSS, we can derive the spectral index distribution. This is presented in Figure~\ref{fig:spixdist}. For situations where multiple AMI-LA sources match a single NVSS source, we have summed the flux densities and used the quadrature sum of the uncertainties. 

We have also placed limits on the spectra of sources not detected in the NVSS by assuming a 1.4~GHz flux density of 2.25 mJy (the $5\sigma$ threshold for point sources in the NVSS). From Figure~\ref{fig:spixdist} our measured spectral index distribution (unfilled histogram) is well-described by a single Gaussian centred on $\alpha_{\rm{high}} = -0.75\pm0.30$.

From Figure~\ref{fig:spixdist} there is a single source with positive spectral index, J102807+675604. This source is unresolved, with a 15.5~GHz flux density of $4.91\pm0.32$~mJy and a two-point spectral index $\alpha=0.36\pm0.07$. However, its counterpart in the NVSS is faint: with a flux density of $2.1\pm0.4$ mJy, it lies below the nominal $5\sigma$ NVSS limit, and the inverted spectrum might otherwise be considered spurious. Nevertheless, cross-matching with the SCG325 catalogue \citep{riseley2016} reveals that this source (SCG325\_J102807+675602) has a 325~MHz flux density of $0.54\pm0.09$ mJy. We will discuss cross-matching with the SCG325 catalogue in the next section.

For sources without NVSS counterparts, the distribution is also centered around approximately $\alpha \simeq -0.8$; however, these are \emph{lower}-limits to the spectral index, and their spectra may be significantly flatter. We note that while we might expect some flux density variability for flat-spectrum objects at 15.5~GHz, our catalogue appears to be dominated by steep-spectrum objects, which tend to exhibit little variability \citep[e.g.][]{waldram2010}. As such, variability should not be a concern for this spectral index study.

\subsubsection{Colour-colour diagram}\label{sec:colourcolour}
We can better model the spectral energy distribution (SED) for sources by cross-matching all three catalogues. Using a cross-matching distance of 20~arcsec, we find unique matches for 39 sources in common across our AMI-LA catalogue, the NVSS and SCG325 catalogues. Many sources in the AMI-LA catalogue have multiple matches at 325~MHz. Where these existed, we took the total flux to be the sum of flux densities and the uncertainty to the be the quadrature sum of the individual uncertainties. 

\begin{figure}
	\begin{center}
		\includegraphics[width=0.47\textwidth]{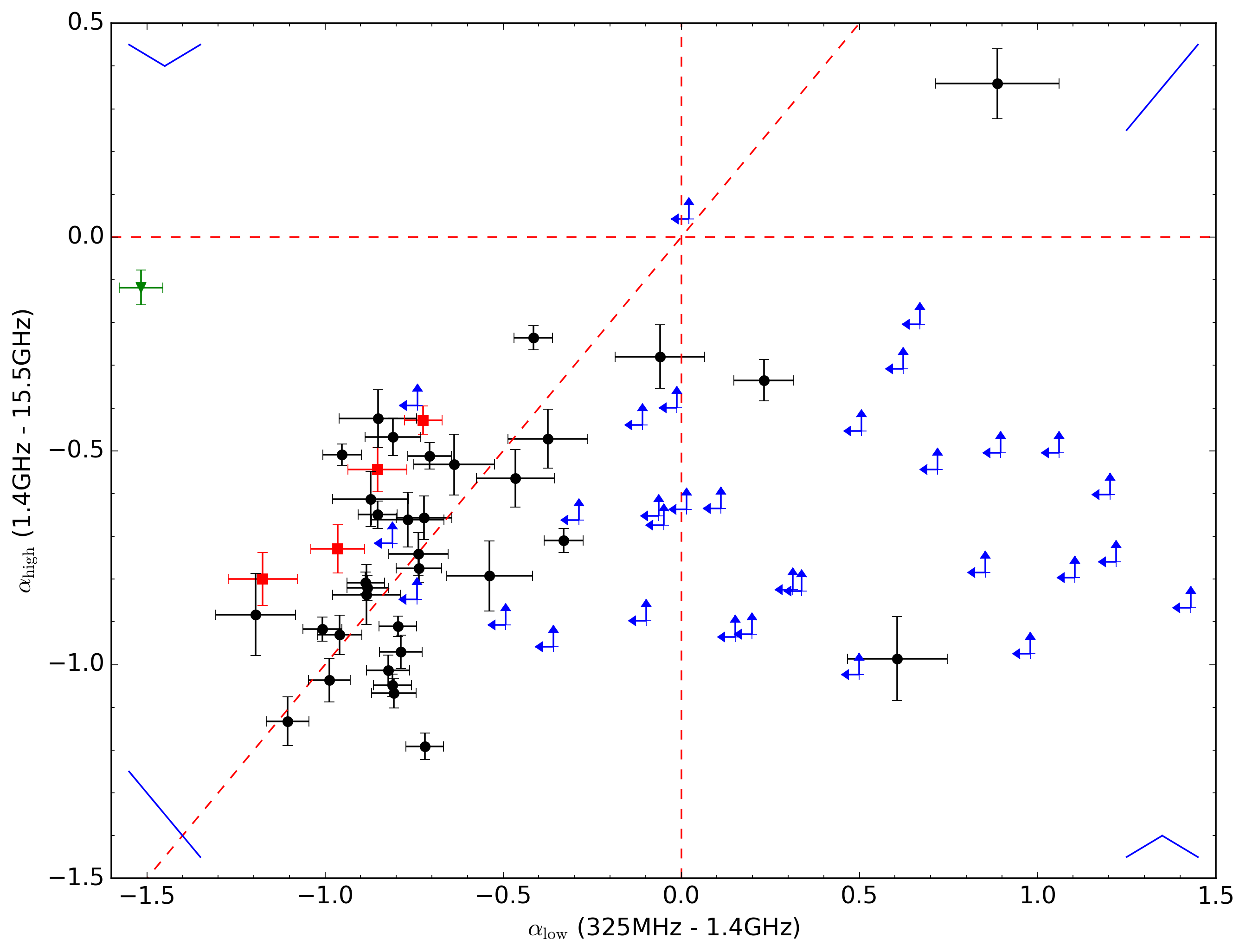} \\
	\end{center}
\caption{Radio colour-colour plot ($\alpha_{\rm{low}}$ against $\alpha_{\rm{high}}$) for sources common to our AMI-LA catalogue, the NVSS and the SCG325 catalogues. $\alpha_{\rm{low}}$ ($\alpha_{\rm{high}}$) is the spectral index calculated between 325~MHz and 1.4~GHz (1.4~GHz and 15.5~GHz). The horizontal and vertical lines indicate a spectral index of zero; the diagonal line indicates $\alpha_{\rm{low}} = \alpha_{\rm{high}}$. The blue lines in the corners of the plot denote the shape of typical spectra for sources in the respective quadrants. Red squares and green triangle points mark sources where the multi-frequency association is more complex (see \S\ref{sec:complex}). Blue limit arrows indicate the location in the colour-colour plot of sources detected at 325~MHz and 15.5~GHz, but not at 1.4~GHz.}
\label{fig:colourcolour}
\end{figure}

While we note the large resolution difference between our AMI-LA catalogue (approx. 50~arcsec) and the SCG325 catalogue (13~arcsec) the GMRT possesses sufficient short baseline coverage to retain sensitivity to sources up to 32~arcmin in extent at 325~MHz; no sources in this region of the SCG325 survey are larger than around 1--2 arcmin in the image plane \citep[see for example Figure 3 of][]{riseley2016}. As such, we expect minimal bias from resolution effects in our colour-colour distribution. We present the radio colour-colour plot (i.e. $\alpha_{\rm{low}}$ vs. $\alpha_{\rm{high}}$) in Figure~\ref{fig:colourcolour}.

From Figure~\ref{fig:colourcolour}, the majority of sources with measured spectra exhibit no spectral break between 325~MHz and 15.5~GHz. However, there are a number of sources that exhibit spectral features significantly different to the majority of the population. We will discuss these sources in this section (postage stamp images at 325 MHz, 1.4 GHz and 15.5 GHz are presented in Figure~\ref{fig:postage}). Red/green square/triangle markers denote sources where there are multiple and/or complex associations between catalogues. 

Additionally, we have plotted the limits for sources detected in the SCG325 catalogue and our AMI-LA catalogue but not in the NVSS catalogue - these are shown with blue limit arrows. These limits suggest a potentially significant population of sources with peaked spectra that are not yet revealed due to the poor sensitivity of the NVSS.

We note that this spectral index distribution accounts for some 53 per cent of sources in our AMI-LA catalogue. At present we are limited by the sensitivity of the NVSS catalogue; crossmatching with our deeper combined e-MERLIN and JVLA catalogue at 1.4~GHz (Battye et al., in prep) will enable us to better understand the spectral index distribution in this field, and provide a tool with which to separate star-forming galaxy and AGN populations. This will be the subject of future work.

\subsubsection{Complex associations}\label{sec:complex}
There are five sources in our AMI-LA catalogue where cross-matching with the SCG325 catalogue is more complex. These are marked in red and green in our colour-colour plot, and we present postage stamps of these sources in Figure~\ref{fig:complex}. For the first four sources -- Panels a) through d) -- multiple counterparts exist in the SCG325 catalogue for a single NVSS/AMI-LA entry, even though some lie beyond the 20~arcsec matching radius. We have indicated the positions of sources we cross-matched between the three catalogues; the resulting three-point spectra are marked by red squares in Figure~\ref{fig:colourcolour}.

For the final source -- panel e) -- multiple components exist in all catalogues. In the SCG325 catalogue, this source comprises three components. In the NVSS catalogue and the AMI-LA catalogue, there are two. The dominant (Western) component exhibits typical synchrotron behaviour at these frequencies ($\alpha_{\rm{low}}=-0.95$ and $\alpha_{\rm{high}}=-0.51$) whereas the Eastern component appears to exhibit a sharp break in the spectrum -- denoted by the green triangle in Figure~\ref{fig:colourcolour}. Given the comparatively low resolution of the NVSS, these components appear blended in the image plane; as such, we suggest that the 1.4~GHz flux density of this second component is underestimated, and its spectrum should perhaps be considered spurious.

\subsubsection{Gigahertz-peaked spectra}
We find two sources (J102012+681236 and J102900+681517; respectively the top and centre panels in Figure~\ref{fig:postage}) that exhibit spectral breaks. J102900+681517 is detected at a high SNR in both the AMI-LA and SCG325 catalogues; in the NVSS it is detected at the $5\sigma$ level. For this source, $\alpha_{\rm{low}}=0.23\pm0.08$ and $\alpha_{\rm{high}}=-0.27\pm0.05$. 

J102012+681236 however, is detected at much lower significance by the NVSS (nominally $3\sigma$) and the AMI-LA (nominally $6\sigma$). The low-/high-frequency spectra are also significantly steeper than that measured for J102900+681517: $\alpha_{\rm{low}}=0.61\pm0.14$ and $\alpha_{\rm{high}}=-0.99\pm0.10$. These sources are natural gigahertz-peaked spectrum (GPS) candidates \citep[e.g.][]{odea1991,odea1998,Callingham2017}. 

Given that $\alpha_{\rm{low}}$ and $\alpha_{\rm{high}}$ for J102900+681517 are not significantly steep, we note that this source could be a flat-spectrum radio source which may exhibit high-frequency variability. Additional monitoring observations would be required to discriminate between these scenarios. No optical classifications are available for these sources, nor is any redshift information available. We also an additional nearby source is seen in the GMRT image of J102900+681517 (see the top row of Figure~\ref{fig:postage}). However, this lies outside the matching radius of 20~arcsec and is assumed to be unrelated.

\subsubsection{High-frequency peaked spectra}
From Figure~\ref{fig:colourcolour}, there is a source (J102807+675604) which shows a steeply inverted low-frequency spectrum $\alpha_{\rm{low}}=0.89\pm0.17$; as discussed in the previous section, this also possesses an inverted high-frequency spectrum $\alpha_{\rm{high}}=0.36\pm0.08$. As such, this source qualifies as a high-frequency-peaked (HFP) source. Given that  $\alpha_{\rm{high}} < \alpha_{\rm{low}}$, it is possible that this source exhibits a turnover near 15.5~GHz; alternatively, this source may exhibit some high-frequency variability.

This source is also compact at the resolution of all instruments considered here (see the bottom row of Figure~\ref{fig:postage}). A potential host (PSO~J102807.164+675602.718) is catalogued at this location in the Pan-STARRS database \citep{Flewelling2016} although no redshift information is available. The deep multi-band optical data taken as part of the SuperCLASS project should enable us to shed further light on all sources identified in our catalogue.

\section{Clustering Analysis}\label{sec:clustering}
Previously, we showed that our catalogue contains a similar number of sources to that expected from the S$^3$ simulation. However, given that our survey area should be dominated by a super-cluster (where we might expect overdensities up to a factor $\sim10$ compared to the field) this is unexpected. For this reason, we have performed a 2D kernel density estimation (KDE) analysis of the source sky position distribution as a further effort to search for clustering effects, using a Gaussian kernel.

\begin{figure}
	\begin{center}
		\includegraphics[width=0.47\textwidth]{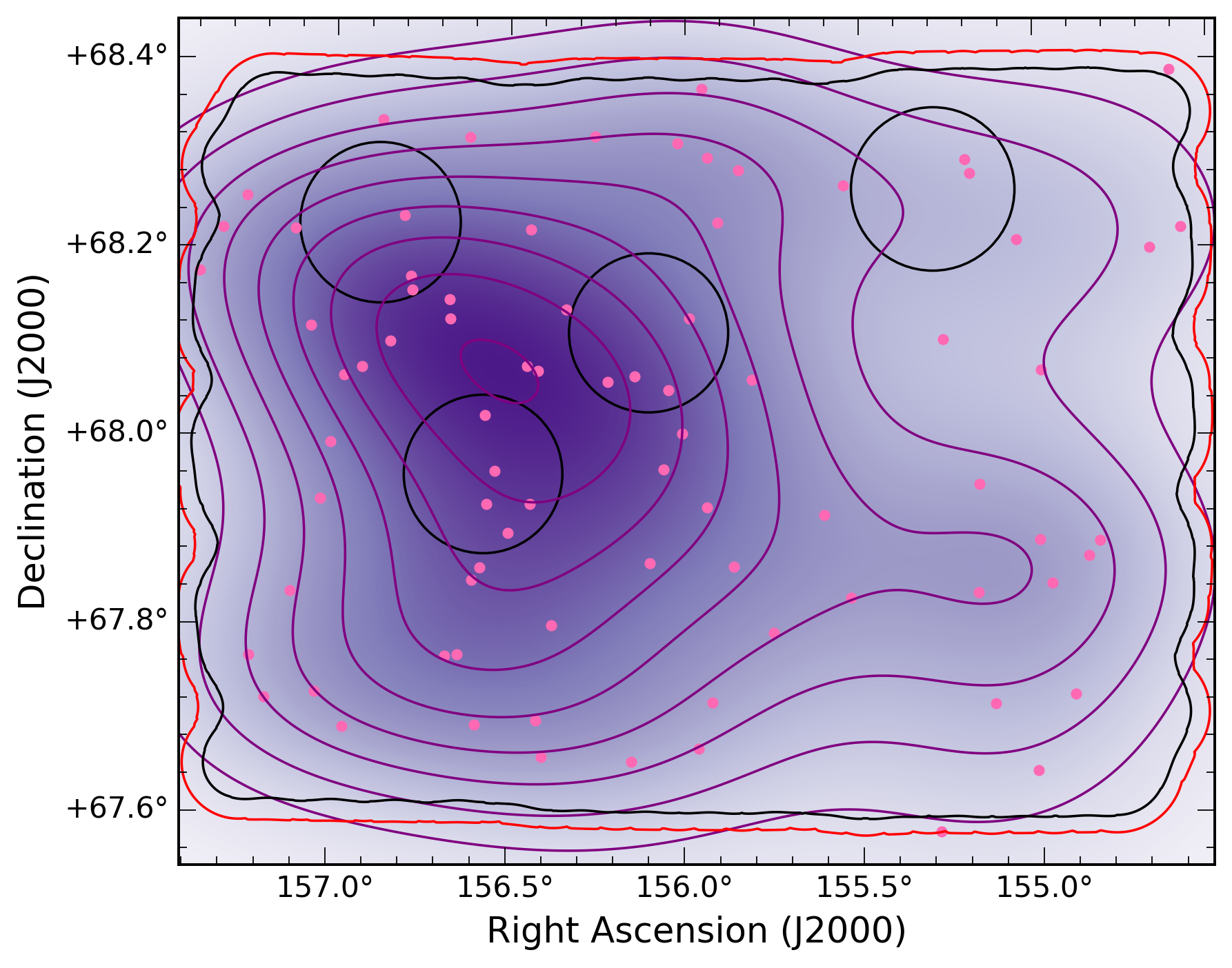} 
		\includegraphics[width=0.47\textwidth]{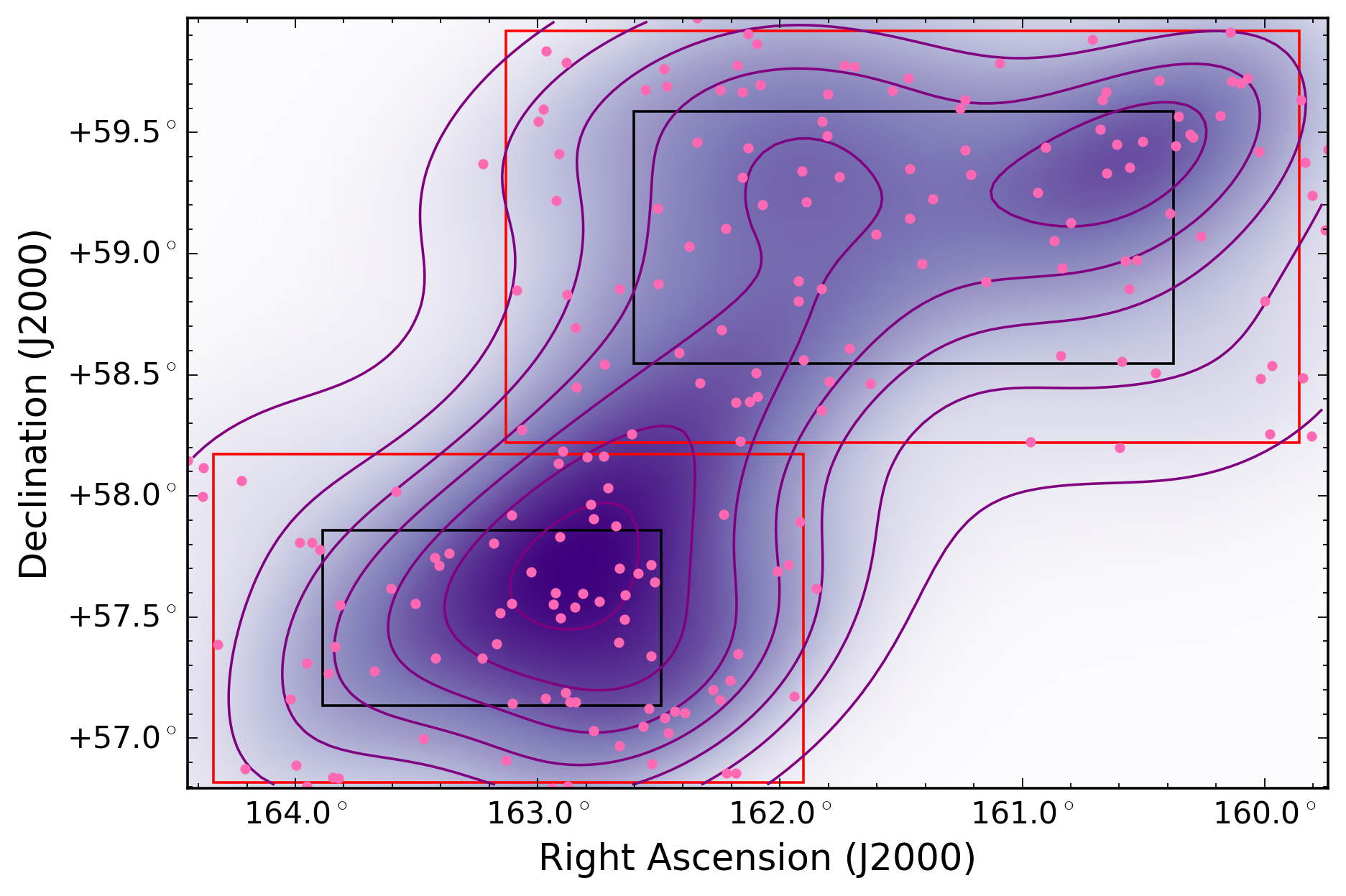} 
	\end{center}
\caption{KDE plots for source positions derived from AMI-LA observations of the SuperCLASS field (top) and Lockman Hole \protect\citep[bottom;][]{whittam2013}. Contours are presented to guide the eye. Source positions are indicated by pink markers. The positions of the clusters in the SuperCLASS super-cluster are overlaid for reference. Black (red) contours denote the regions complete to 0.5~mJy (1~mJy) based on the noise map in Figure~\ref{fig:rms} (for the SuperCLASS field) and the regions listed in \protect\cite{davies2011} (for the Lockman Hole).}
\label{fig:kdeplot}
\end{figure}

We also performed a 2D KDE analysis of the AMI-LA catalogue selected from 10C observations of the Lockman Hole \citep{whittam2013}. Given the differences in target choice between the SuperCLASS field and the Lockman Hole, we might expect to find evidence of clustering in our field, but not the Lockman Hole. The KDE plots are presented in Figure~\ref{fig:kdeplot}, where we have overlaid the source catalogues for reference.

From Figure~\ref{fig:kdeplot}, both plots appear to exhibit evidence of clustering. However, the Lockman Hole survey depth (overlaid in Figure~\ref{fig:kdeplot}) suggests that the apparent clustering is related to the variation in sensitivity across the field. For the SuperCLASS field, however, the survey depth is generally consistent across the field (as shown by the contours; see also Figure~\ref{fig:rms}) and almost the entire survey area is complete to 0.5~mJy. As such, the apparent clustering exhibited in the top panel of Figure~\ref{fig:kdeplot} is likely to be real, suggesting that the majority of radio sources detected by the AMI-LA may be associated with the super-cluster. However, without host identification or redshift measurement (which will be provided by the deep optical data taken as part of the SuperCLASS project) this cannot yet be confirmed.

\section{Source Count}\label{sec:src}

\subsection{Construction of Source Count}\label{sec:construction}
We constructed the 15.5~GHz source count by binning our catalogue according to integrated flux density -- using the \emph{peak} flux density for unresolved sources, as described in \S\ref{sec:unresolved}. We established our bins using a width of 0.15~dex before adjusting the bins in an attempt to maintain a minimum of 9 sources per bin. We were largely successful; aside from the highest flux density bin, the minimum number of sources in any given bin is 7, while a typical bin has $\sim8$ sources. We note that, due to the relatively small survey area, our final flux density bin spans a broad flux density range ($8.84-80$~mJy).

We derive the corrected number of sources in each bin $(N_{\rm{c}})$ by weighting each source according to the reciprocal of its visibility area $(V_{\rm{area}})$. The visibility area is derived from the rms map (Figure~\ref{fig:rms}) and is presented in Figure~\ref{fig:visarea}. Therefore the number of sources in a given bin is
\begin{equation}
	N_c = \sum_{\rm{nsrc}} \left( V_{\rm{area}} \right)^{-1} .
\end{equation}
The raw $(N)$ and corrected $(N_{\rm{c}})$ number of sources are presented in Table~\ref{tab:src_counts}, along with the differential source count $({\rm{d}}N/{\rm{d}}S)$ and the Euclidean-normalised differential source count $({\rm{d}}N/{\rm{d}}S ~ S_{\rm{c}}^{2.5})$, where we have used the geometric mean flux density $(S_{\rm{c}})$ for each bin to perform the normalisation. We quote Poisson errors $(\sqrt{N})$ on the Euclidean-normalised differential source count. Note that we have excluded the lowest flux density bin $(0.190-0.239~{\rm{mJy}})$ from our source count analysis, as approximately 50 per cent of sources in this bin had a low visibility area (less than 10 per cent) and as such may not be representative of the whole field.

\begin{table}
\begin{center}
\caption{Differential source count for the SuperCLASS survey region at 15.5~GHz. Columns list the flux density range covered, the central flux density $(S_c)$, raw $(N)$ and corrected $(N_c)$ number of sources in each bin, as well as the differential source count $({\rm{d}}N/{\rm{d}}S)$ and the Euclidean-normalised source count $({\rm{d}}N/{\rm{d}}S ~ S_c^{2.5}$)}
\label{tab:src_counts}
\begin{tabular}{cccccc}
\hline
& & & & & \\
Flux bin &	$ S_{\rm{c}} $ & $N$ &  $N_c$ & $\frac{{\rm{d}}N}{{\rm{d}}S}$ & $\frac{{\rm{d}}N}{{\rm{d}}S} S_c^{2.5}$ \\
$[$mJy$]$ & $[$mJy$]$ & & & $[$Jy$^{-1}$ sr$^{-1}]$ & $[$Jy$^{1.5}$ sr$^{-1}]$\\
\hline
$ 0.239 - 0.301 $ & 0.268 & 9 & 20.82 & $ 1.17 \times10^9 $ & $ 1.37 \pm 0.30 $ \\
$ 0.301 - 0.379 $ & 0.338 & 8 & 13.10 & $ 5.83 \times10^8 $ & $ 1.22 \pm 0.34 $ \\
$ 0.379 - 0.450 $ & 0.413 & 7 & 9.18 & $ 4.49 \times10^8 $ & $ 1.56 \pm 0.51 $ \\
$ 0.450 - 0.577 $ & 0.51 & 8 & 9.72 & $ 2.65 \times10^8 $ & $ 1.56 \pm 0.50 $ \\
$ 0.577 - 0.795 $ & 0.677 & 10 & 10.59 & $ 1.69 \times10^8 $ & $ 2.01 \pm 0.62 $ \\
$ 0.795 - 0.950 $ & 0.869 & 7 & 7.19 & $ 1.61 \times10^8 $ & $ 3.58 \pm 1.34 $ \\
$ 0.950 - 1.50 $ & 1.194 & 11 & 11.05 & $ 6.97 \times10^7 $ & $ 3.43 \pm 1.03 $ \\
$ 1.50 - 3.00 $ & 2.121 & 9 & 9.0 & $ 2.08 \times10^7 $ & $ 4.32 \pm 1.44 $ \\
$ 3.00 - 8.84 $ & 5.15 & 7 & 7.0 & $ 4.16 \times10^6 $ & $ 7.92 \pm 2.99 $ \\
$ 8.84 - 80.0 $ & 26.59 & 3 & 3.0 & $ 1.46 \times10^5 $ & $ 16.87 \pm 9.74 $ \\
\hline
\end{tabular}
\end{center}
\end{table}

\subsection{Cosmic Variance}
\cite{heywood2013} investigate the effects of cosmic variance on differential source count at 1.4~GHz by comparing observations with a number of independently-selected samples from S$^3$. They subsequently derive a mechanism by which the source count uncertainty due to cosmic variance can be estimated for a given radio survey. 

Our fields are much smaller than those considered by \cite{heywood2013} -- from Figure~\ref{fig:rms} and Figure~\ref{fig:visarea}, sources in our faintest flux density bin (which have a typical flux density of 0.268 mJy) can be detected across approximately 40 per cent of our survey area, or $0.356$ square degrees. Based on Figure~2 of \cite{heywood2013} we might expect an uncertainty of approximately 20 per cent for this bin. This uncertainty decreases with increasing flux density, as brighter sources are detectable across a greater fraction of our survey area.

\subsection{Bias Effects}
There are two principal effects which may influence our source count distribution, which must be considered: Eddington bias and resolution bias.

\subsubsection{Resolution Bias}
Sources are catalogued based on their \emph{peak} flux density (compared to the local rms), whereas the contribution to source count is determined according to \emph{integrated} flux density. For high-resolution surveys, it can be the case that resolved sources -- which would otherwise contribute to the counts -- can fall below the required signal-to-noise threshold to be detected. 

We previously used the formalism of \cite{hales2014a} to model the effect of resolution bias on our GMRT survey at 13 arcsec resolution, where 37 per cent of sources in our catalogue were resolved. We found that resolution bias was around the five per cent level at 10 mJy, and increasing toward faint flux densities \citep{riseley2016}. 

As discussed in \S\ref{sec:unresolved}, we have treated all sources in our catalogue as being point-like. With a minimum baseline of 18~m (corresponding to $\sim800\lambda$) the maximum scale size detectable by the AMI-LA is of the order of 260 arcsec. This is approximately a factor three higher than the largest \emph{convolved} source image-plane major axis. In light of this -- and our use of natural weighting -- we do not expect that resolution bias will play a significant role.

We also note that we have not merged the double sources in our catalogue for the purposes of deriving the source count. This is for a number of reasons: firstly, double sources comprise some 15 per cent of our catalogue. Secondly, the relatively small number statistics used to derive the source count will dominate the uncertainty. Additionally, however, we cannot conclusively say that the two components of these double sources are associated, as optical host galaxies cannot yet be identified.

\begin{figure*}
	\centering
	\includegraphics[width=0.8\textwidth]{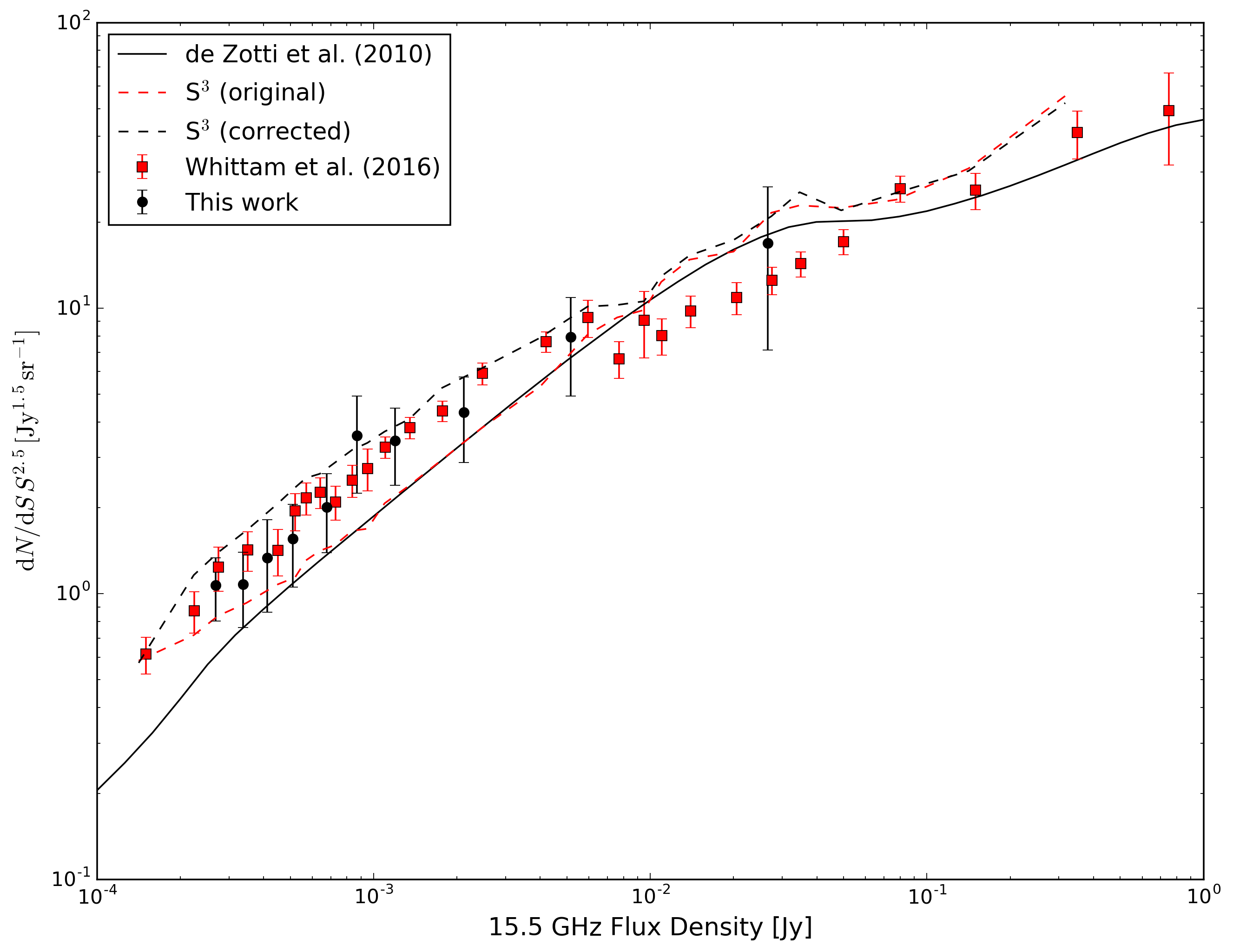}
	\caption{Euclidean-normalised differential source count at 15.5~GHz from this work (black filled circles) and the combined counts from the 9C, 10C, and ultra-deep 10C surveys \protect\cite[red squares;][and references therein]{whittam2016}. The solid line marks the \protect\cite{zotti2010} AGN source count model. The dashed red curve denotes the source count at 18~GHz from S$^3$ below 0.5~Jy; dashed black curve denotes the S$^3$ source count corrected for an enhanced core fraction -- see \protect\cite{Whittam2017}.}
	\label{fig:src_counts}
\end{figure*}

\subsubsection{Eddington Bias}
Eddington bias is caused by fluctuations in image noise that scatter faint sources to higher flux densities; given that the number counts decline rapidly with increasing flux density, this is more likely than the opposite effect. We would expect this to be most significant toward the flux density limit of our survey. Previous surveys with the AMI-LA found that Eddington bias effects were at the $\sim7$ per cent level \citep{davies2011} and closely balanced by incompleteness corrections. \cite{whittam2016} do not apply a correction for this reason.

Our previous consideration of resolution and Eddington bias at 325~MHz indicate that these effects shift the source count in opposite directions -- accounting for resolution bias provides a boost to the differential source count, whereas the correction for Eddington bias provides a negative shift. We will not apply a correction for Eddington bias.

\subsection{Source Count Profile}
We present the Euclidean-normalised differential source count from Table~\ref{tab:src_counts} in Figure~\ref{fig:src_counts}, along with deep differential source count at 15.7~GHz from the 9C/10C surveys and the Ultra-deep 10C extension \citep[][and references therein]{whittam2016}. For reference, we also overlay the 15~GHz \cite{zotti2010} AGN model\footnote{Available at \url{http://w1.ira.inaf.it/rstools/srccnt/srccnt_tables.html}.}. We make no attempt to correct for the frequency difference between these source count.

From Figure~\ref{fig:src_counts}, it is clear that our source count agree well with those from the literature, although the smaller number of sources in our catalogue means the uncertainties on our source count are greater than those of \cite{whittam2016}. Both the AGN model from \cite{zotti2010} and the S$^3$ simulation under-predict the faint source count by around a factor $\sim2$; whereas our results broadly agree with the `corrected' S$^3$ source count \citep{Whittam2017}. This discrepancy at faint flux densities could reasonably be attributed to the low core fraction previously identified from other surveys at these frequencies \cite[e.g.][]{mahony2011,whittam2016}. Toward higher flux densities, where there is better agreement between S$^3$ and the \cite{zotti2010} model, our results are also consistent.

Whilst the contribution from star-forming galaxies is clearly detected in the low-frequency source count from this field \citep{riseley2016} the 15.5~GHz source count profile does not exhibit evidence of this population. Ultra-deep studies at lower frequencies suggest that the population of SFG is smaller than predicted by simulations \citep[e.g.][although note the authors' comment on the high resolution of their observations]{guidetti2017}; the deep multi-wavelength data taken as part of the SuperCLASS project may shed some light on the contribution of star-forming galaxies to the observed source population.

\section{Conclusions}\label{sec:conc}
We have presented the results of deep 15.5~GHz AMI-LA observations of the SuperCLASS field, a galaxy supercluster known to contain five Abell clusters at redshift $z\sim0.2$. We achieved a nominal sensitivity of $32.0~\umu$Jy beam$^{-1}$ toward the field centre, with a typical sensitivity better than $60~\umu$Jy beam$^{-1}$ across the majority of the survey area, which covers approximately 0.9 square degrees. We compile a source catalogue above $5\sigma_{\rm{local}}$, which contains 80 sources.

We have derived the radio colour-colour distribution for sources common to three previous radio surveys covering this field. From this distribution, we identify three sources which exhibit spectral index trends different to the majority of the population. One of these we identify as a high-frequency-peaked spectrum source; two we identify as gigahertz-peaked-spectrum sources. 

Our investigation of the spectral index distribution is hindered by the shallow survey limit of the NVSS. As such, our catalogue appears to be dominated by steep-spectrum sources; however, due to the shallow NVSS limit, faint flat-spectrum objects detected by the AMI-LA will be undetectable at 1.4~GHz. The deeper e-MERLIN and JVLA data taken as part of this project will allow us to better investigate this spectral index distribution.

We use this catalogue to derive the differential source count at 15.5~GHz, down to a limiting flux density of 0.24 mJy. Our source count are broadly consistent with those from previous ultra-deep observations of other fields; they do not show any evidence of an emerging population of star-forming galaxies that has been seen at lower frequencies, or is predicted by models. In addition, models of the AGN population are shown to underpredict the differential source count toward fainter flux densities; it has been suggested that this is due to the flat-spectrum cores of radio galaxies contributing more to the observed source count than current models predict. 

\subsection{Future work}
This work forms the initial part of a broader AMI survey campaign on this field. We also have quasi-simultaneous observations with the AMI Small Array (AMI-SA) in hand. With sensitivity to emission on much larger angular scales, the AMI-SA has proven successful at detecting the Sunyaev-Zel'dovich (SZ) effect from galaxy clusters (for example  \citealt{Perrott2015}; \citetalias{Shimwell2013}). We will use the catalogue produced in this work to subtract the compact source population from the field to investigate the mass distribution of the super-cluster via the SZ effect in a future work.

\subsection{Acknowledgements}
We thank the engineers and staff at the Mullard Radio Astronomical Observatory for maintenance and operation of AMI, which is supported by Cambridge and Oxford Universities. We are grateful for IT knowledge exchange with the SKA project. The AMI telescope, AMS \& TC gratefully acknowledge support from the European Research Council under grant ERC-2012-StG-307215 LODESTONE. We thank our anonymous referee for constructive comments that have improved our manuscript. 

CJR acknowledges an OCE Fellowship from CSIRO. IH, MLB, and BT are supported by a European Research Council Starting Grant (grant number 280127). MLB is an STFC Advanced/Halliday fellow (grant number ST/I005129/1). YCP acknowledges support from a Trinity College Junior Research Fellowship.

This research has made use of data from the OVRO 40-m monitoring program (Richards, J. L. et al. 2011, ApJS, 194, 29) which is supported in part by NASA grants NNX08AW31G, NNX11A043G, and NNX14AQ89G and NSF grants AST-0808050 and AST-1109911. This work has also made use of the NASA/IPAC Extragalactic Database (NED), operated by JPL under contract with NASA, as well as NASA's Astrophysics Data System (ADS).

\bibliographystyle{mnras}
\bibliography{SuperCLASS_AMI}

\appendix

\section{Images of sources discussed in text}
\begin{figure*}
	\centering
	\includegraphics[width=0.25\textwidth]{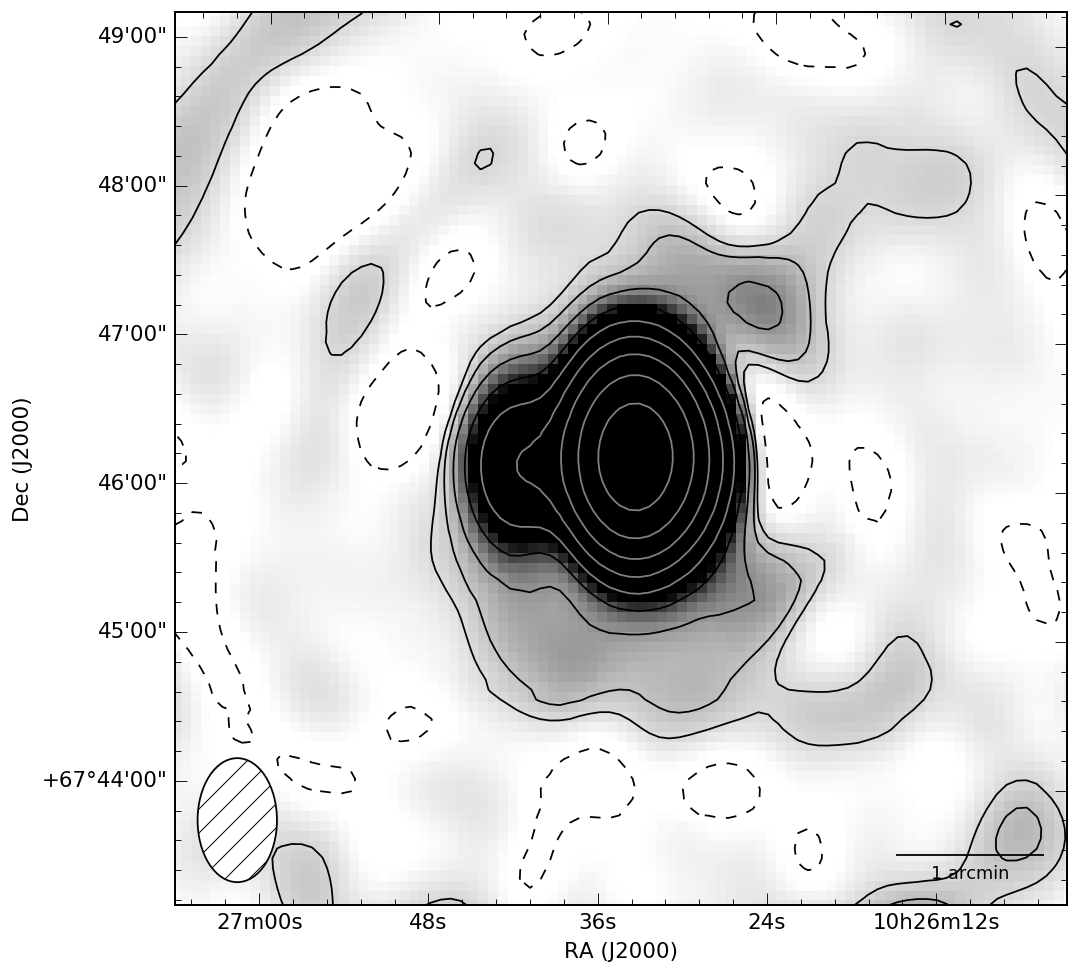} 
	\includegraphics[width=0.26\textwidth]{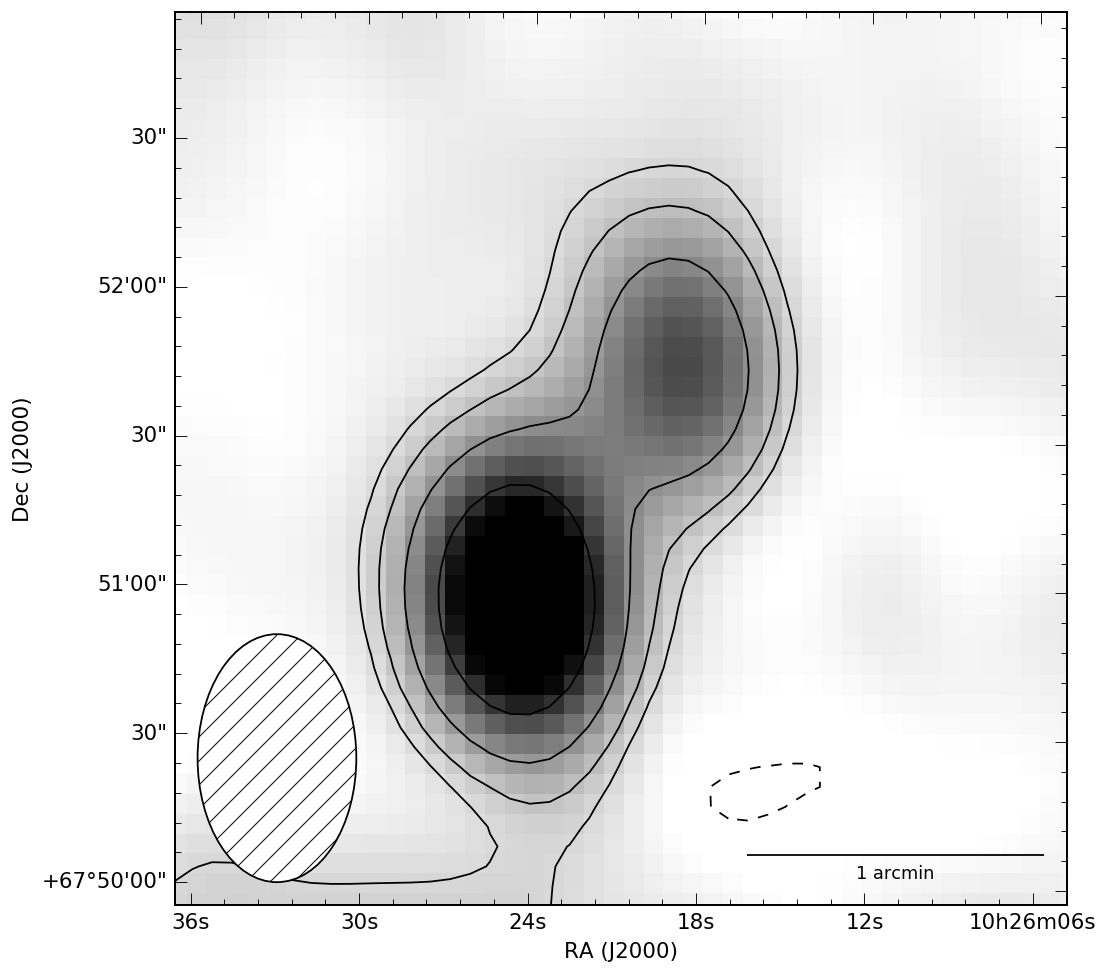} 
	\includegraphics[width=0.25\textwidth]{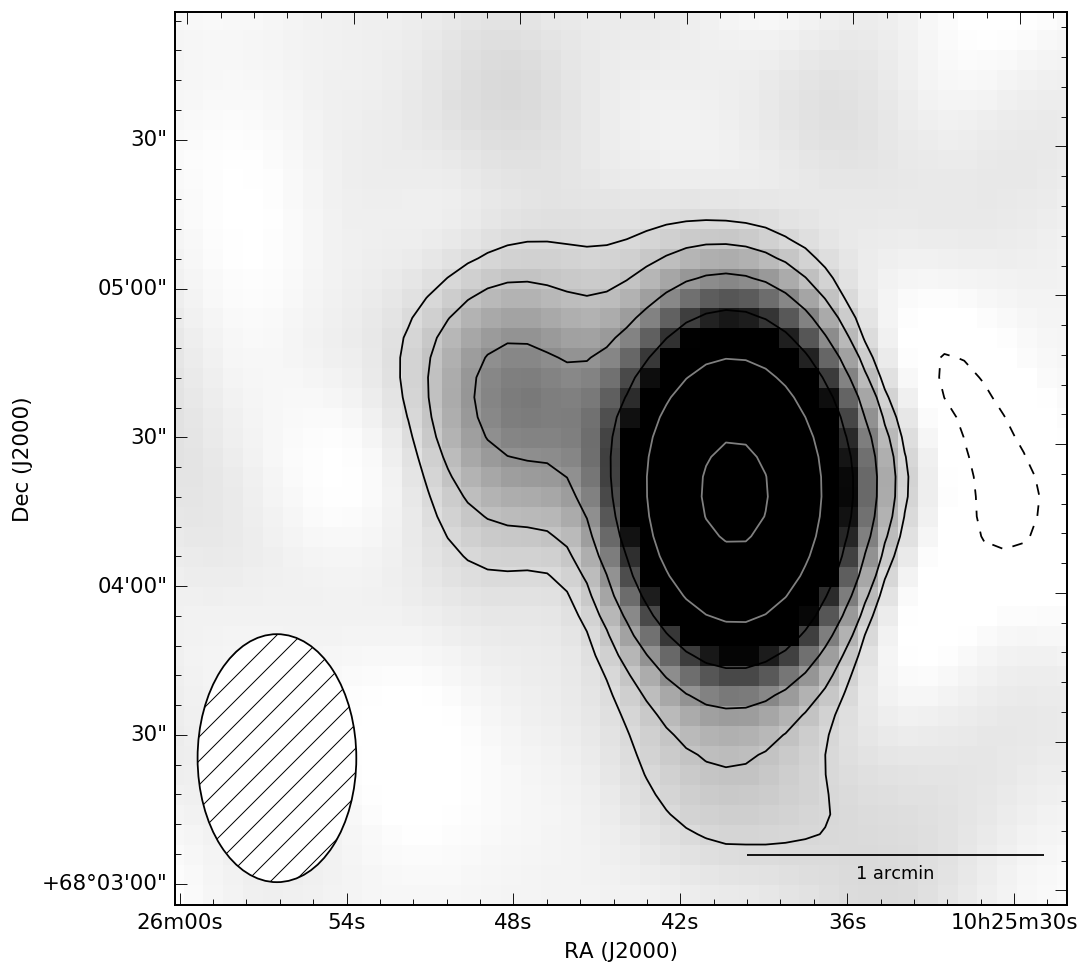} \\
	\includegraphics[width=0.25\textwidth]{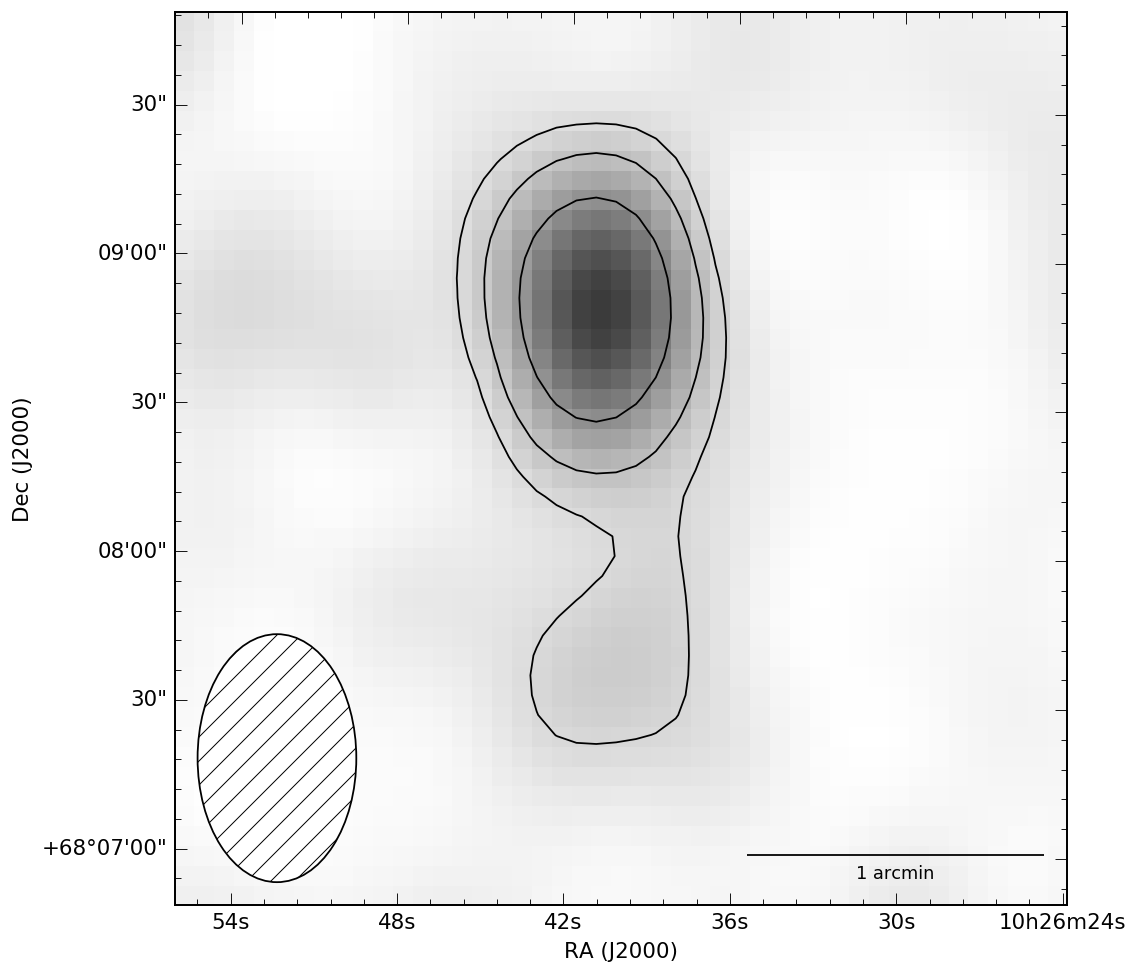}
	\includegraphics[width=0.25\textwidth]{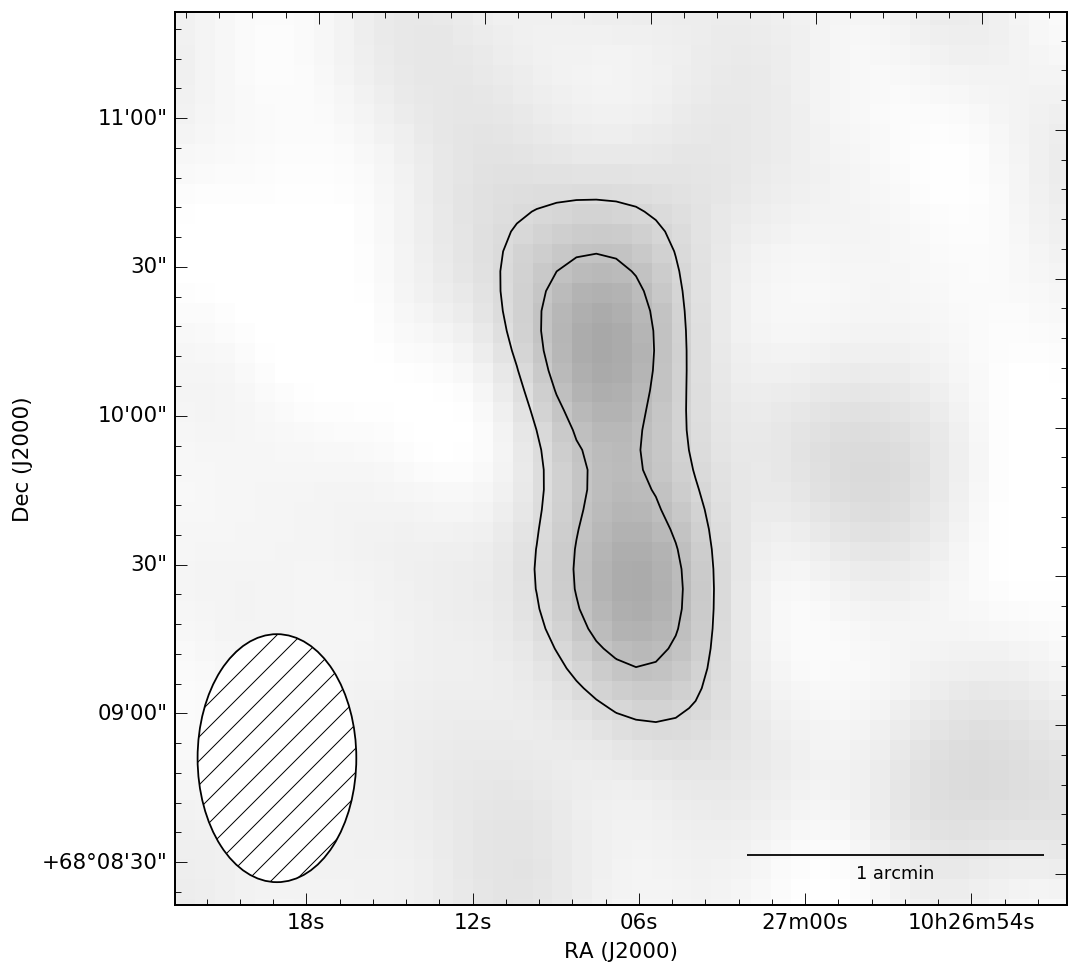} 
	\includegraphics[width=0.25\textwidth]{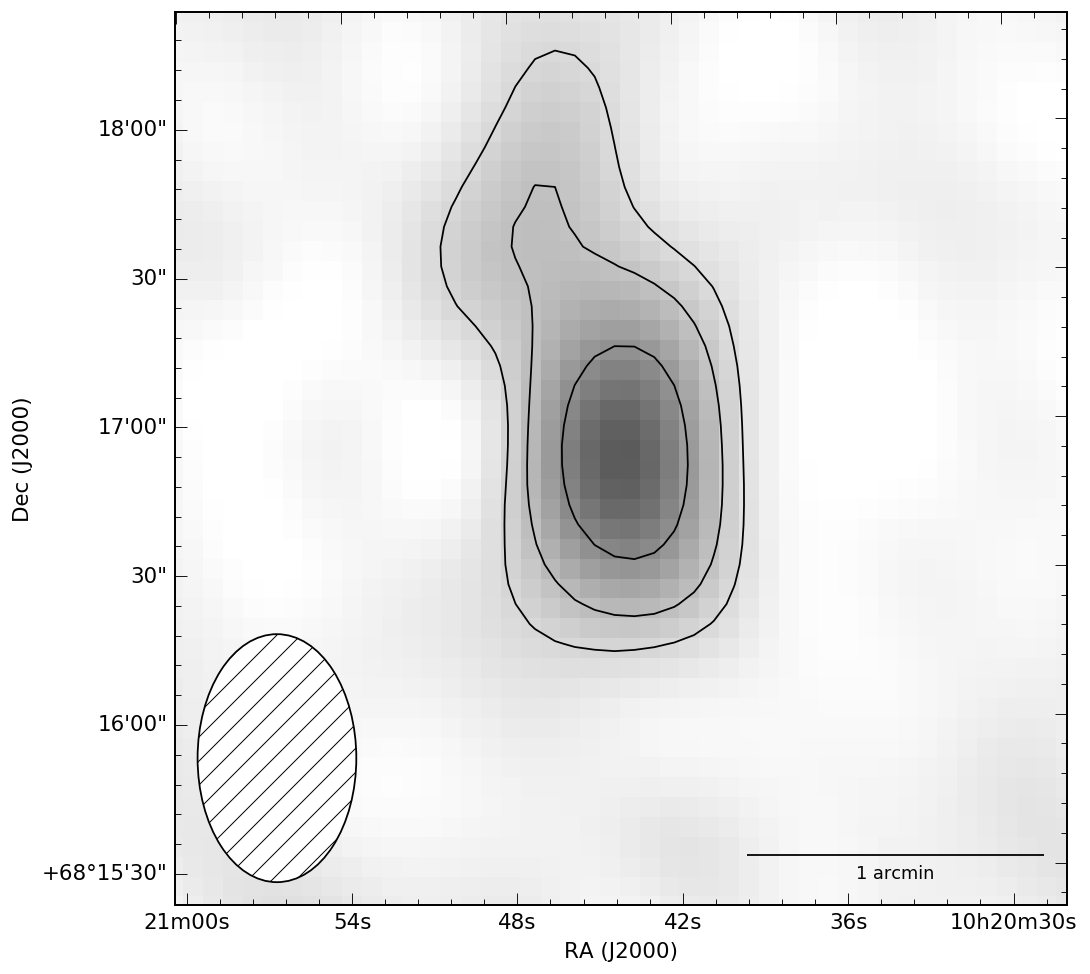} 
\caption{Postage stamp images of sources classified as doubles in our catalogue. Solid contours start at $5\sigma_{\rm{nom}}$ and scale by a factor two, where $\sigma_{\rm{nom}}=32.0~\umu$Jy beam$^{-1}$ is the nominal off-source noise; the $-5\sigma_{\rm{nom}}$ level is denoted by the dashed contour. The hatched ellipse denotes the AMI-LA beam area. The limited dynamic range around the bright double source (top-left panel) is clearly visible. All panels are set to matching colour scales, which saturate at $50\sigma_{\rm{nom}}$.}
\label{fig:doubles}
\end{figure*}

Figure~\ref{fig:doubles} presents postage stamp images of double sources taken with the AMI-LA.

\begin{figure*}
	\centering
	\includegraphics[width=0.7\textwidth]{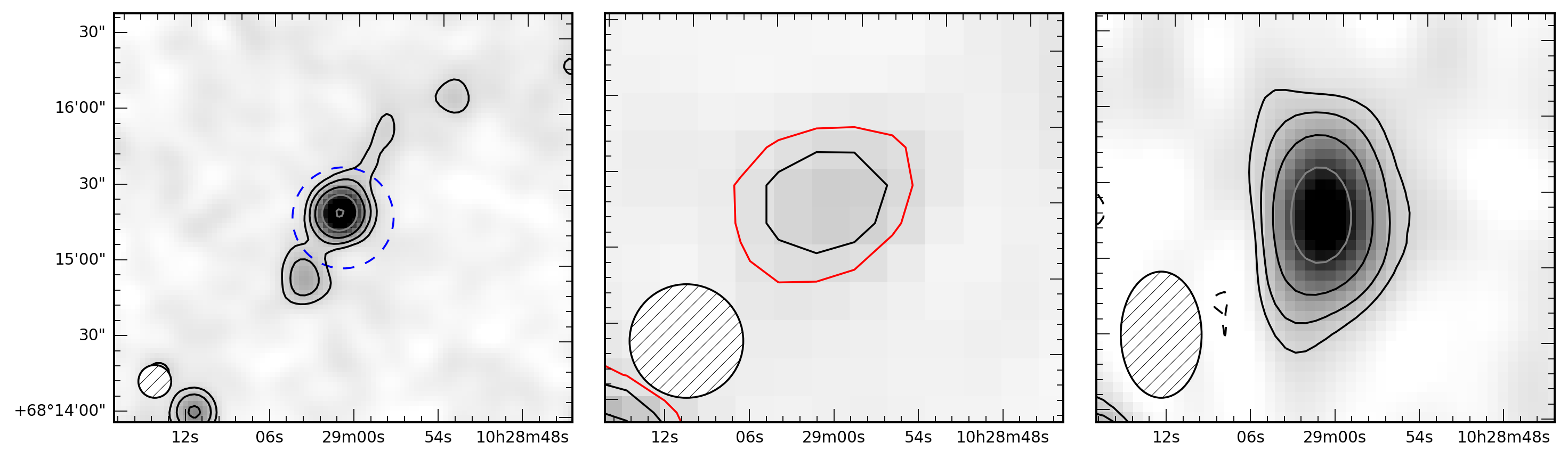} \\
	\includegraphics[width=0.7\textwidth]{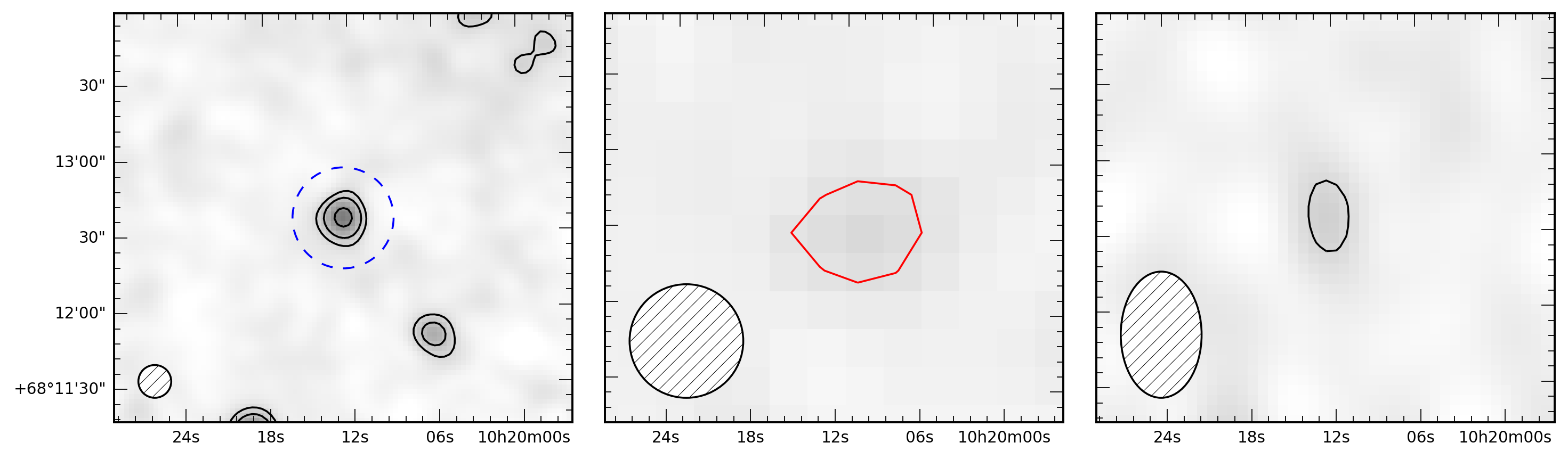} \\
	\includegraphics[width=0.7\textwidth]{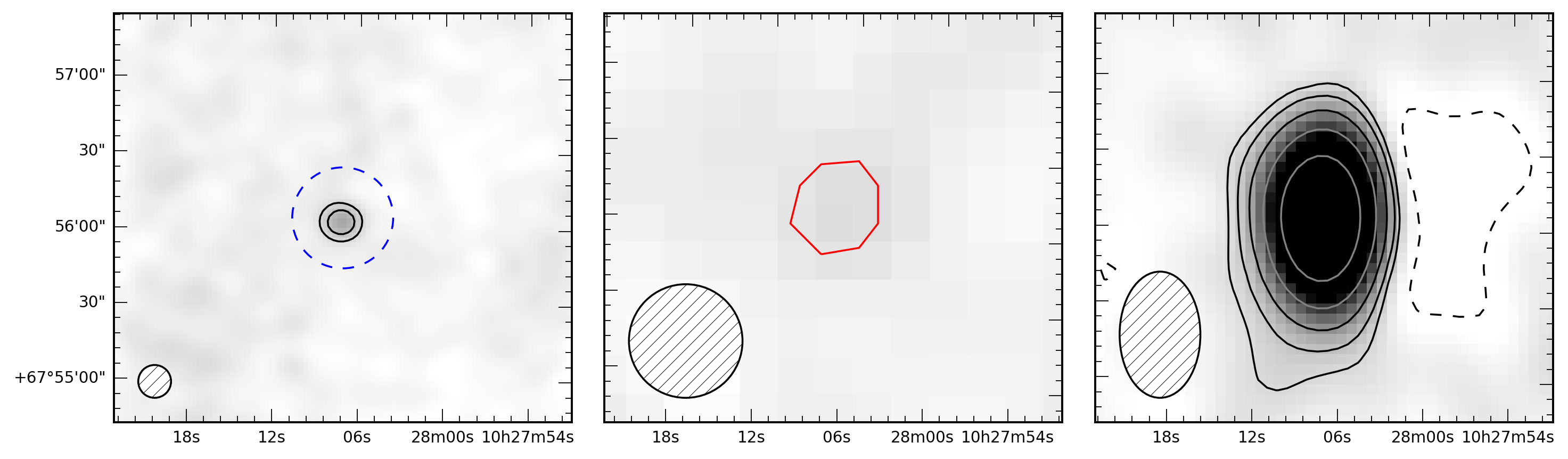}
\caption{Postage stamp images of sources with peaked or rising spectra, identified in \S\ref{sec:colourcolour}. Left / centre / right panels present GMRT / NVSS / AMI-LA images. All colourscales range from $-3\sigma_{\rm{nom}}$ to $50\sigma_{\rm{nom}}$ where the nominal off-source noise $(\sigma_{\rm{nom}})$ is $34~/~450~/~31~\umu$Jy beam$^{-1}$. The beam area of each survey is indicated by the hatched ellipse in the lower-left corner. Black/gray contours start at $5\sigma_{\rm{nom}}$ and scale by a factor two; the red contour denotes the $3\sigma_{\rm{nom}}$ NVSS level. Sources are J102900+681517, J102012+681236 and J102807+675604, from top to bottom. The dashed blue circle denotes a 20~arcsec radius centred on the NVSS coordinates for each source; this was the radius within with sources in the GMRT image were considered to match the higher-frequency data.}
\label{fig:postage}
\end{figure*}

Figure~\ref{fig:postage} presents postage stamp images of sources with peaked spectra from GMRT, NVSS and AMI-LA observations of this field \citep[respectively][and this work]{riseley2016,condon1998}, identified from the colour-colour plot presented in \S\ref{sec:colourcolour} and are further discussed there.

\begin{figure*}
	\centering
	\begin{tabular}{c}
	\includegraphics[width=0.88\textwidth]{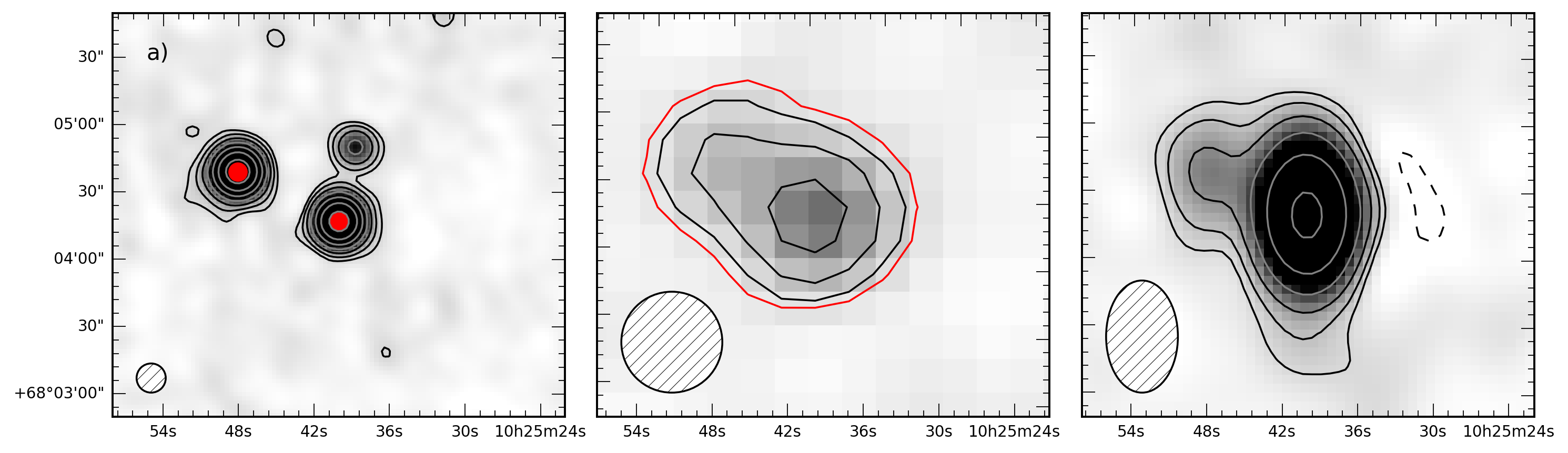} \\
	\includegraphics[width=0.88\textwidth]{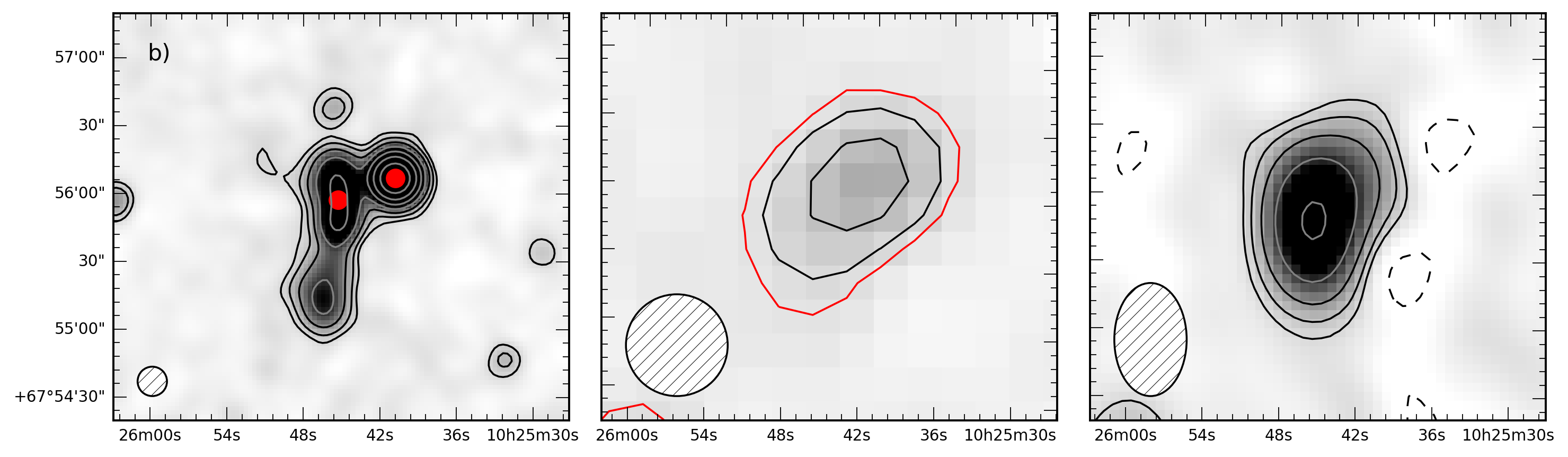} \\
	\includegraphics[width=0.88\textwidth]{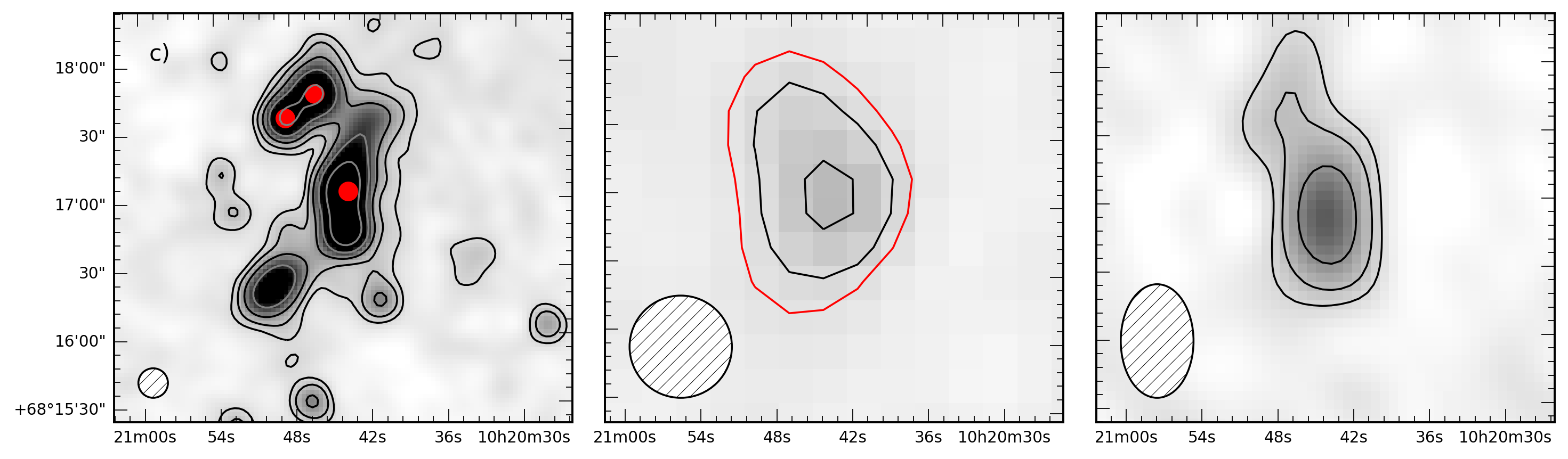} \\
	\includegraphics[width=0.88\textwidth]{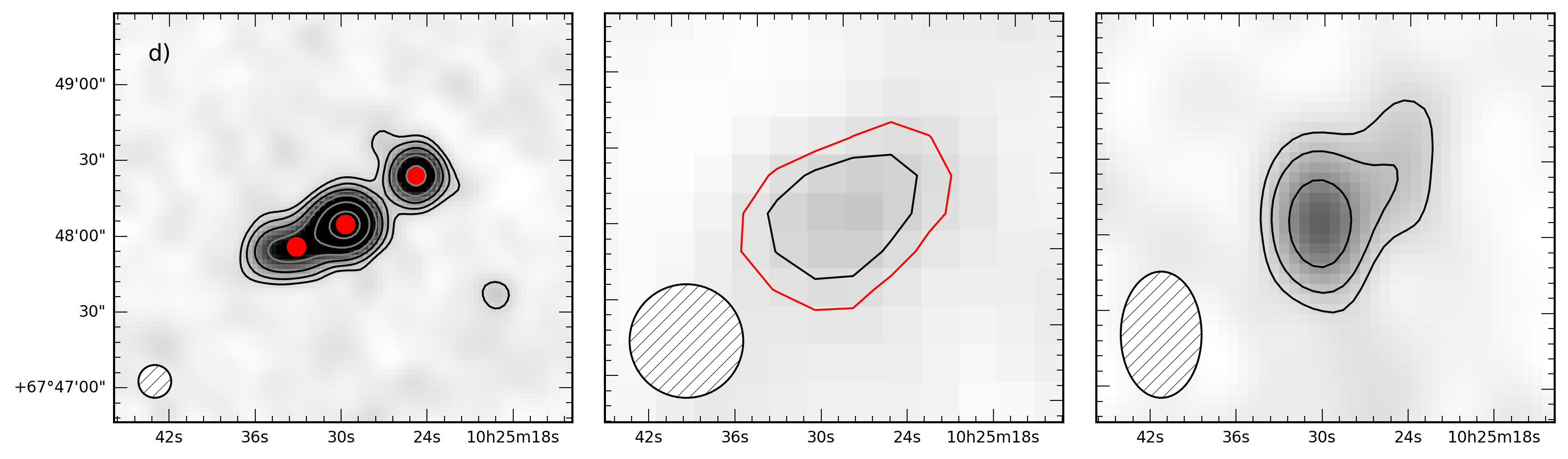} \\
	\includegraphics[width=0.88\textwidth]{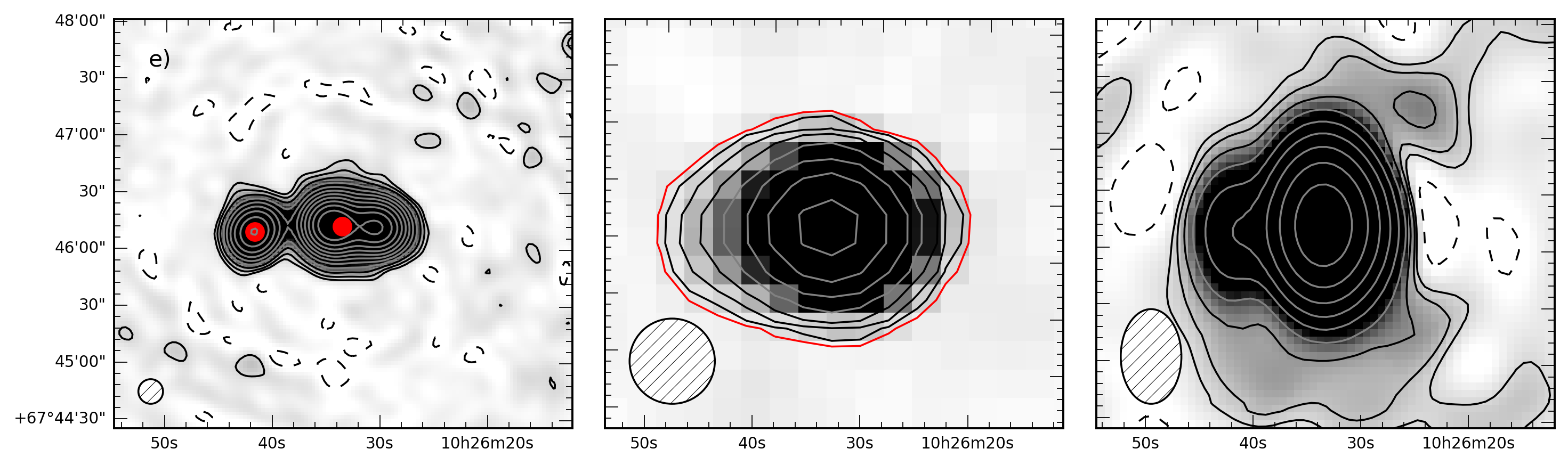} \\
	\end{tabular}
\caption{Sources common to the SCG / NVSS / AMI-LA catalogues (respectively left / centre / right panels) which are identified as complex or having multiple matches in Table~\ref{tab:src_cat}. The colour scales and contour levels are as per Figure~\ref{fig:postage}. Red markers denote the position of components at 325~MHz which are deemed to match the higher-frequency data and are used to derive the colour-colour plot in Figure~\ref{fig:colourcolour}.}
\label{fig:complex}
\end{figure*}

Figure~\ref{fig:complex} presents images of sources identified from our colour-colour analysis where multiple and/or complex matches exist across the three catalogues considered in this work.

\end{document}